\documentclass[]{aa}  

\usepackage{graphicx}
\usepackage{revsymb}        
\usepackage{amsmath}        
\usepackage[usenames]{color}
\usepackage{natbib}
\graphicspath{{figures/}}
\usepackage[normalem]{ulem}
\usepackage{cancel}

\begin{document}   

\title{Pulsations of rapidly rotating stars}

   \subtitle{I. The ACOR numerical code}

   \author{R-M. Ouazzani\inst{1,2}, M-A. Dupret\inst{2}, D. R. Reese\inst{2} }

          \institute{LESIA, UMR8109, Universit\'e Pierre et Marie Curie, Universit\'e Denis Diderot, Observatoire de Paris, 92195 Meudon Cedex, France\\
            \email{rhita-maria.ouazzani@obspm.fr}
            \and
            Institut d'Astrophysique et de G\'eophysique de l'Universit\'e de Li\`ege, All\'ee du 6 Ao\^ut 17, 4000 Li\`ege, Belgium   }
          
          \date{Received September 15, 1996; accepted March 16, 1997}

%

 
  \abstract
   {Very high precision seismic space missions such as CoRoT and Kepler provide the means of testing the modeling of transport processes in stellar interiors. For some stars, such as solar-like and red giant stars, a rotational splitting is measured. However, in order to fully exploit these splittings and constrain the rotation profile, one needs to be able to calculate them accurately. For some other stars, such as $\delta$ Scuti and Be stars, for instance, the observed pulsation spectra are modified by rotation to such an extent that a perturbative treatment of the effects of rotation is no longer valid.}
   {We present here a new two-dimensional non-perturbative code, called ACOR (\textit{Adiabatic Code of Oscillation including Rotation}) which allows us to compute adiabatic non-radial pulsations of rotating stars, without making any assumptions on the sphericity of the star, the fluid properties ({\it i.e.} baroclinicity) or the rotation profile.}
   {The 2D non-perturbative calculations fully take into account the centrifugal distortion of the star and include the full influence of the Coriolis acceleration. The numerical method is based on a spectral approach for the angular part of the modes, and a fourth-order finite differences approach for the radial part.}
   {We test and evaluate the accuracy of the calculations by comparing them with those coming from TOP (\textit{Two-dimensional Oscillation Program}) for the same polytropic models. We illustrate the effects of rapid rotation on stellar pulsations through the phenomenon of avoided crossings.}
   {As shown by the comparison with TOP for simple models, the code is stable, and gives accurate results up to near-critical \rm rotation rates. }

   \maketitle

\color{black}
\section{Introduction}
Rotation plays a key role in the evolution of stars across the Hertzsprung-Russell  diagram (which shows the distribution of stars in the luminosity versus effective temperature plane). For instance, the centrifugal distortion breaks the thermal equilibrium, and provokes large scale currents called meridional circulation  \citep{Eddington1925}. These currents as well as shear and baroclinic instabilities modify the angular momentum distribution \citep{Zahn1992,Mathis2004}, and thereby the rotation profile inside the stars. Meanwhile, these processes transport chemical elements, and change the evolution of the stars.
That is the reason why determining rotational profiles inside stars is crucial for modeling stellar structure and evolution.

Asteroseismology is currently the only tool that allows such a determination. But rotation also changes stellar pulsations. The centrifugal force distorts the resonant cavity of the pulsations, and the Coriolis force modifies the modes' dynamic.
Usually, for slow rotators, the effects of rotation are taken into account as a perturbation of pulsations (see for instance \citealt{Ledoux1951} for first-order effects, \citealt{Gough1990} and \citealt{Dziembowski1992} for second-order effects, and \citealt{Soufi1998} for third-order effects). But these methods, although elegant and simple of use have shown their limits \citep{Reese2006,Suarez2010}. 
On the one hand these perturbative methods approximate the effects of the centrifugal force on acoustic pulsations in stars that show very high surface velocities, such as $\delta$ Scuti and Be stars. On the other hand, they approximate the impact of the Coriolis force on gravity modes in stars whose surface rotates slowly, but in which the pulsation periods are of the same order as their rotation period ($P_{rot}\sim P_{osc}$), such as SPB stars or $\gamma$ Doradus. Finally, perturbative methods may also fail in modeling pulsations of cooler stars, such as sub-giant or red giant stars, with very low surface velocities but rapidly rotating cores, and in which all pulsation modes are of a mixed nature, i.e. gravity close to the core and acoustic in the envelope \citep[see for example ][]{Beck2012}.

This concerns many stars in  CoRoT and Kepler observation fields. Hence, if one wants to correctly extract the rotation profile from seismic observations, they need to correctly apprehend the effects of rotation on pulsations. 

 This work aims at presenting a model which accurately takes into account the non-perturbative effects of rotation on oscillation spectra. 
A new two dimensional non-perturbative code is presented. The 2D non-perturbative calculations fully take into account the centrifugal distortion of the star, and include the full influence of the Coriolis acceleration. This 2D non-perturbative code is useful for studying pulsational spectra of highly distorted evolved models of stars, as well as stars presenting highly differential rotation profiles. 
Section \ref{S2} introduces the basic pulsation equations in spheroidal geometry using a coordinate system adapted to the star's shape. Section \ref{S3} is dedicated to the numerical method which is based on a spectral approach for the angular part of the modes, and a finite difference method, which is accurate to fourth order, for the radial part. In Sect. \ref{S4}, we test these calculations and evaluate their accuracy. Finally, in Sect. \ref{S5}, the results are validated by comparing them with those of \cite{Reese2006}, and an example of application is given in Sect. \ref{S6}.

\section{Basic equations in spheroidal geometry}
\label{S2}
When dealing with the subject of computing the pulsations of rotating stars, one has to face two main  issues. Firstly, rotation, through the centrifugal force, distorts the resonant cavity of the pulsations. If a solenoidal rotation profile is assumed (around a north-south axis), the azimuthal symmetry is conserved whereas the spherical one is broken: the star takes on a spheroidal geometry. This centrifugal distortion, if it is strong enough has to be treated by a two-dimensional approach.
Secondly, when rotation is accounted for in the pulsation equations,  the Coriolis acceleration enters the momentum equation and modifies the dynamics of the modes. If small enough, this Coriolis effect can be approximated by perturbative methods. But for moderate to high rotational velocities, as well as for high order g modes when $P_{rot}\sim P_{osc}$, the perturbative treatment is no longer relevant, and a non perturbative approach has to be adopted. 

\subsection{Spheroidal geometry}
\label{S2s1}

\begin{figure} [t!]
\begin{center}
\hspace*{-1cm}\includegraphics[scale=0.33, angle=-90]{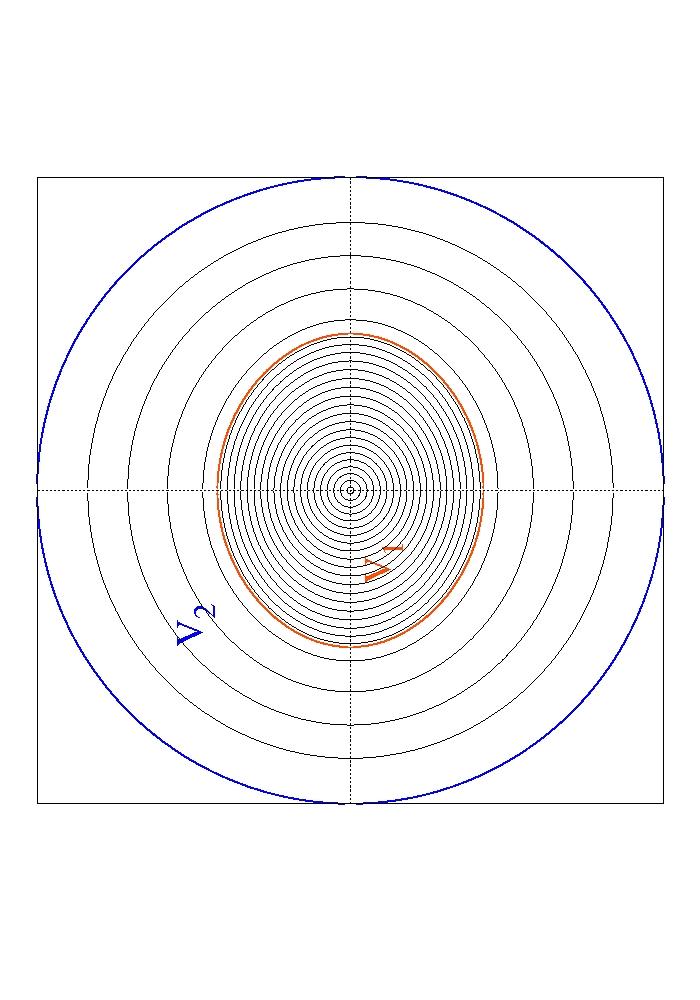}
\caption{\label{Coord_syst_polyt}Coordinate system used in ACOR. V$_1$ extends from the center to the stellar surface, and $V_2$ encompasses the star.}
\end{center}
\end{figure}

Given the distorted shape of a rotating star, we choose a new coordinate system which defines the star's surface at a constant pseudo-radial coordinate.
To do so, we adopt a multidomain approach, which consists in dividing the physical space into domains whose boundaries correspond to the model's discontinuities \citep[$\mathrm{e.g.}$ ][]{Canuto1988}: one domain V$_1$, which encloses the star, and one external domain, V$_2$, bounded by the stellar surface and a sphere of twice the equatorial radius. Following \cite{Bonazzola1998}, we introduce a coordinate system which goes from spherical symmetry at the center to a spheroidal shape at the stellar surface, and back to a spherical geometry at the external boundary of V$_2$.
In this system, the radial coordinate, $\zeta$, is no longer $r$, the distance to the center. However, for a fixed colatitude $\theta$, $r$ and $\zeta$ are related thanks to the following continuous and bijective function:\\  
In domain $V_1$:
\begin{align}
r(\zeta,\theta) \, &= \, (1-\epsilon) \, \zeta \, + \, \frac{5\zeta^3 \, - \, 3\zeta^5}{2} \left( R_s(\theta) - 1 + \epsilon \right)\, ,
\end{align}
where $\zeta$ ranges from 0 to 1, $\epsilon=1-R_{pol}/R_{eq}$ is the flatness,  and $r(\zeta=1,\theta)=Rs(\theta)$ the stellar surface.\\
In domain $V_2$: 
\begin{align}
r(\zeta,\theta) \, =  \, 2& \, \epsilon \, + \, (1-\epsilon) \, \zeta \, \nonumber \\
\hspace*{0.5cm}+& \, \left( 2\zeta^3 - 9\zeta^2+12 \zeta -4 \right) \, \left( R_s(\theta) - 1 + \epsilon \right) \, ,
\end{align}
where $\zeta$ ranges from 1 to 2, $\zeta=2$ corresponding to a spherical surface which encompasses the star (see Fig. \ref{Coord_syst_polyt}). 
The use of such a coordinate system helps significantly with establishing the boundary conditions.

\noindent Then one has to define a new vector basis. We choose the following orthogonal spheroidal basis:
\begin{align}
\overrightarrow{a_{\zeta}}\, &= \, \frac{1}{1+\frac{r_{\theta}^2}{r^2}} \, \left( \overrightarrow{e_{r}}\, - \, \frac{r_{\theta}}{r} \, \overrightarrow{e_{\theta}} \right)\, ,  \nonumber \\
\overrightarrow{a_{\theta}}\, &= \, \frac{1}{1+\frac{r_{\theta}^2}{r^2}} \, \left( \overrightarrow{e_{\theta}}\, + \, \frac{r_{\theta}}{r} \, \overrightarrow{e_{r}} \right)\, ,   \\
\overrightarrow{a_{\varphi}}\, &= \, \overrightarrow{e_{\varphi}}\, , \nonumber
\end{align}
where
\begin{itemize}
\item[$\bullet$] $\overrightarrow{a_{\theta}}$ is tangential to iso-$\zeta$ surfaces in the meridional plane,
\item[$\bullet$]$\overrightarrow{a_{\zeta}}$ is directly orthogonal to $\overrightarrow{a_{\theta}}$ in the meridional plane,
\item[$\bullet$]$\overrightarrow{a_{\varphi}} \, = \, \overrightarrow{e_{\varphi}}$,
\end{itemize}
and where $r_{\theta} = \partial_{\theta} r$.

In a forthcoming paper, this method will be applied to a realistic stellar model using three domains: one which encloses the convective core, the second the radiative envelope, and the third an external domain, which allows us to avoid discontinuities at the top of the convective zone. 

\subsection{Basic equations}

\label{ss_basic-eq}
Here, we consider the stellar interior to be an inviscid self-gravitating fluid, where the magnetic field is neglected. Therefore its physics is governed by the conservation of mass, momentum and energy, the energy transfer equation, and Poisson's equation for the gravitational potential \citep[see for example][]{{Kippenhahn1994}}. 

Using the Eulerian formalism, we compute the oscillation modes as the adiabatic response of the structure to small perturbations -- $\mathrm{i.e.}$ of the density, pressure, gravitational potential and velocity field. As in \cite{Unno1989}, we perturb the hydrodynamics equations around the equilibrium state. 
Considering that the velocity field in the equilibrium state is only due to rotation, the linearized equations in the inertial frame are given by
\begin{eqnarray}
\label{conservation_pert_xi}  \partial_t \rho' \, + \, \mathbf{\nabla} \cdot \left( \rho_0 \mathbf{v'}\, + \, \rho' \mathbf{v_0}\right) \, = \, 0 \, , \hspace*{1.5cm}\\
\left[\left(\partial_t + \Omega \partial_{\varphi}\right) v'_{i}\right] \mathbf{e_{i}}+ 2 \mathbf{\Omega} \times \mathbf{v'} + \left(\mathbf{v' . \nabla \Omega} \right) r \sin\theta \mathbf{e_{\varphi}} \, = \, \nonumber \\
\hspace*{1.5cm}- \frac{1}{\rho_{0}} \mathbf{\nabla}p' - \mathbf{\nabla} \Phi' + \frac{\rho'}{\rho_{0}^{2}} \mathbf{\nabla}p_{0}\, , \hspace*{1.8cm}\\
\label{relat_ad_pert} \left( \partial_t + \mathbf{v_0} \cdot \mathbf{\nabla} \right)  \left( \frac{\rho'}{\rho_{0}} - \frac{p'}{\Gamma_{1} p_{0}} \right) + \mathbf{v}' \cdot \left(\frac{\mathbf{\nabla} \rho_0}{\rho_0} - \frac{\mathbf{\nabla} p_{0}}{\Gamma_{1} p_0} \right)\, = \, 0\, , \\
\mathbf{\nabla}^2 \Phi' \, = \, 4 \pi G \rho' \, ,  \hspace*{2.5cm}
\end{eqnarray}
where primed quantities denote Eulerian perturbations, whereas quantities with the subscript ``$_0$'' correspond to the stationary state. The symbol $\mathbf{e_{i}}$ stands for the spherical basis vectors. Note that the energy conservation equation has been replaced by the adiabatic relation.
\noindent Given that the equilibrium state is stationary and axisymmetric, the time and azimuthal dependences are of the form $e^{i(\omega t+m \varphi)}$, where $\omega$ is the pulsation frequency, and $m$ the azimuthal order of the pulsation mode.

We put the system of equations in non-dimensional form using the following transformations:
\begin{align}
\tilde{\rho} = \left( \frac{4 \pi R_{eq}^3}{M} \right) \rho   \hspace{0.7cm}   \tilde{p} = \left( \frac{4 \pi R_{eq}^4}{G M^2} \right)p& \hspace{0.7cm} \tilde{\phi} = \left( \frac{R_{eq}}{GM} \right) \phi \nonumber   \\
\hspace*{-0.5cm}\sigma = \frac{\omega}{\Omega_k}  \hspace{1.5cm}&  \tilde{\Omega}= \frac{ \Omega}{\Omega_k} \nonumber
\end{align}
where $\Omega_k$ stands for the Keplerian critical velocity, {\it i.e.} the rotation velocity at which the centrifugal force compensates gravity at the equator.

From now on, we work in terms of non-dimensional variables and drop the tilded notation.

Rather than projecting the motion equation onto the basis vectors, we choose to decompose it into one radial, one poloidal and one toroidal field. This decomposition allows us to obtain separate partial differential equations for the radial and angular coordinates, and helps to reduce the number of unknowns, as will be shown later.

Moreover, we apply the change of variable $\pi' \, = \, p' \, / \, \rho$ so as to avoid singularity problems at the surface of polytropic models, and we introduce an auxiliary variable $\rm d \Phi'$, as well as the equation relating it to $\Phi'$: $\rm d\Phi'=\partial \Phi / \partial \zeta = \partial_{\zeta} \Phi'$, thereby reducing Poisson's equation from a second order differential equation to two first order ones.

\subsection{Spectral expansion}
\label{s2ss3}
Thanks to an appropriate choice of the coordinate system, the behavior of the eigenfunctions on iso-$\zeta$ surfaces is smooth. Therefore, the angular behavior of pulsation modes is well described in terms of an expansion on the basis of spherical harmonics. Such a spectral method is known to be very well suited in fluid dynamics. For instance, for a function of class $\mathcal{C}^{\infty}$, the numerical error decreases as $e^{-aM}$, where M is the number of spherical harmonics in the expansion and $a$ a constant \citep{Canuto1988}.

Therefore, we perform a spectral expansion of the unknowns in terms of the spherical harmonics $Y_{\ell}^{m}$ \citep{Rieutord1987}. Any vector field can be decomposed into a radial, a poloidal and a toroidal part. Therefore, the components of the pulsation velocity field are expressed as follows in the vector basis $(\mathbf{a_{\zeta}},\mathbf{a_{\theta}},\mathbf{e_{\varphi}})$

\begin{align}
  \label{eq_decomp_vit1}
  v'_{\zeta}&= { i} \sum_{\ell \geq \mid m \mid}^{+ \infty} u_{\ell}(\zeta)   Y_{\ell,m}(\theta,\varphi) \, ,\\
  v'_{\theta}&= { i} \sum_{\ell \geq \mid m \mid}^{+ \infty} \left( v_{\ell}(\zeta) \, \partial_{\theta} Y_{\ell,m}(\theta,\varphi) \, + \, w_{\ell}(\zeta) \frac{m}{\sin \theta}  Y_{\ell,m}(\theta,\varphi) \right)\, ,\nonumber  \\
  v'_{\varphi}&= - \sum_{\ell \geq \mid m \mid}^{+ \infty} \left( v_{\ell}(\zeta) \frac{m}{\sin \theta} Y_{\ell,m}(\theta,\varphi) \, + \, w_{\ell}(\zeta) \, \partial_{\theta} Y_{\ell,m}(\theta,\varphi) \right)\, .\nonumber 
  \label{eq_decomp_vit2}
\end{align}
All other scalar unknowns, namely $\Phi'$, $d\Phi'$, and $\rho'$, are expanded in the same way as the scaled pressure perturbation:
\begin{align}
  \pi'= \sum_{\ell \geq \mid m \mid}^{+ \infty} \pi'_{\ell}(\zeta)   Y_{\ell,m}(\theta,\varphi) 
\end{align}

\subsection{Symmetries and mode classification}

Because of the symmetries of the equilibrium model with respect to the rotation axis and to the equator, one obtains a separate eigenvalue problem for each value of the azimuthal order, $m$, and each parity, $par$, with respect to the equator. Thus, for a given value of $m$, there are two independent sets of ODEs coupling the spectral coefficients, one with $\ell$ of the same parity as $m$, and the other with opposite parities.
That means, that for a given $m$, when including $M$ terms in the spectral expansion, $\forall j \in [ 1, M ],$ 
\begin{align}
\label{def_l}
&\ell \, = \, \mid m \mid \, + \, 2(j-1) \, + \, par, \, \hbox{ for } \, \pi'_{\ell},\, \phi'_{\ell}, \,{\rm d}\phi'_{\ell}, \,u_{\ell}, \,v_{\ell}, \rho'_{\ell}  \, , \\
\label{def_lp}
&\ell_p \, = \, \mid m \mid \, + \, 2(j-1) \, +\, 1 \, - \, par \hspace{0.2cm} \hbox{ for } \hspace{0.2cm} w_{\ell_{ p}}, 
\end{align}
 with $par=0$ if $\ell$ is of the same parity as $m$ (even mode), and $par=1$ otherwise (odd mode). 
We then obtain a system of ordinary differential equations (ODE) of the variable $\zeta$ for the coefficients of the spherical harmonic expansion $u_{\ell}, \, v_{\ell}, \, w_{\ell_p}, \, \pi'_{\ell}, \, \rho'_{\ell}, \, \Phi'_{\ell}$, and $d\Phi'_{\ell}$.

\subsection{Projections}
After having expanded the unknowns on the basis of spherical harmonics, the second step consists in projecting the equation system onto the spherical harmonics basis which is truncated to $M$ terms, as is the spectral expansion.  
In this subsection, the unknowns are generically designated by $\rm X_i(\zeta,\theta,\varphi)$. Each partial differential equation of the form
\begin{align}
\rm E\left( X_i(\zeta,\theta,\varphi) \right)\, = \, \rm E\left( \sum_{j=1}^{\infty} X_{i,\ell_2}(\zeta)\, Y_{\ell_{2},m}(\theta,\varphi) \right)\, = \, 0
\end{align}
is replaced by a system of $M$ differential equations in $\zeta$, obtained by projecting these equations on a basis of $M$ spherical harmonics:
\begin{align}
\label{eq_principe_decomp} 
\int \frac{\sin \theta \rm d\theta \rm d\varphi}{4 \pi} \, \rm E\left( \sum_{j=1}^{\infty} X_{i,\ell_2}(\zeta)\, Y_{\ell_{2},m}(\theta,\varphi)\right) \,  Y_{\ell_{1},m}^* \, = \, 0\, ,
\end{align}
in which $\ell_1$ and $\ell_2$ take on the values defined in Eqs.(\ref{def_l}) or (\ref{def_lp}). 

The equilibrium quantities, which are expressed as functions of $\zeta$ and $\theta$, are implicitly contained in the operator E.

The impact of the basis dimension (M) on the precision of the computations will be discussed in detail in Sect. \ref{ss_conv}.

\subsection{Boundary conditions}

\label{bound_cond}
In order to complete the eigenvalue problem, it is necessary to specify a number of boundary conditions. The system of equations contains 4  sets of  first order ODEs. Thus, we impose 4 boundary conditions. As the system is solved simultaneously for all layers, two boundary conditions are imposed at the center, one at the stellar surface and one on the external spherical surface of $V_2$ (see Fig. \ref{Coord_syst_polyt}).

Taking boundary conditions at the center is a delicate problem because of the coordinate system. It consists in imposing the regularity of the solutions at the center. To do so, we take the limits of the equations as $\zeta$ goes to zero. The different scalar unknowns behave as follows
\begin{eqnarray}
\pi'_{\ell} \, = \, \mathcal{O} \left( \zeta^{\ell} \right) \hspace{0.5cm} \Phi'_{\ell} \, = \, \mathcal{O} \left( \zeta^{\ell} \right) \hspace{0.5cm} \rho'_{\ell} \, = \, \mathcal{O} \left( \zeta^{\ell} \right) \nonumber
\end{eqnarray}
and the components of the velocity field obey
\begin{align}
u_{\ell} \, = \, \mathcal{O} \left( \zeta^{\ell-1} \right) \hspace{0.3cm} &\hbox{and}  \hspace{0.3cm} u_{0} \, = \, \mathcal{O} \left( \zeta \right),\nonumber \\
v_{\ell} = \mathcal{O} \left( \zeta^{\ell-1} \right)  \hspace{0.3cm} &\hbox{and}  \hspace{0.3cm} v_{0} \, = \, 0 \, ,\nonumber \\ 
w_{\ell} \, = \, &\mathcal{O} \left( \zeta^{\ell} \right) \nonumber 
\end{align}
This results in two algebraic relations between the unknowns. 
A detailed explanation of these central boundary conditions is presented in Appendix \ref{app_cond-cent}.

At the stellar surface, a stress free condition is imposed: $\delta p=0$. The stellar surface is assumed to be on an isobar, at $\zeta=1$, thus the boundary condition corresponds to: 
\begin{align}
i (m\Omega+\sigma) \, \pi' \, = \, - \, \frac{\partial_{\zeta} p_0}{\rho_0} \, v'_{\zeta} \, ,
\end{align}
where the subscript ``$_0$'' stands for equilibrium quantities.

The external condition for the gravitational potential consists in imposing that it vanishes at infinity. This can be achieved by matching the gravitational potential to a vacuum potential at $\zeta=2$, $\mathrm{i.e.}$ on the spherical external surface $V_2$. The advantage in using a spherical boundary is that the Laplace equation is separable for solutions of
the form $X_i(r,\theta,\varphi) = X_i(r) Y_{\ell}^m(\theta,\varphi)$. This gives very simple conditions for a continuous match at the $\zeta=2$ spherical boundary. Specifically, for $\zeta \ge 2$, this equation is decomposed over the spherical harmonic basis, and each component solved separately.  For each $\ell$, this yields two independent solutions: $\Phi'_{\ell} = A\, r^{\ell}$ which diverges at infinity and is therefore discarded, and $\Phi'_{\ell} = B\,/\, r^{(\ell+1)}$ which vanishes. Hence, the corresponding boundary condition is
\begin{align}
\rm d\Phi'_{\ell} \, = \, -\, (\ell+1) \, \frac{r_{\zeta}}{r} \, \Phi'_{\ell}\, ,
\end{align}
where $r_{\zeta} = \partial_{\zeta} r $.

\section{Numerical method}
\label{S3}
The spectral form of equations (\ref{App_eq_Mouvement}), (\ref{App_eq_Divergence}), (\ref{App_rot}), (\ref{App_eq_relat_adiab2}), (\ref{App_eq_Continuite}), (\ref{App_eq-aux}), and (\ref{App_eq_Poisson2}) constitute a first order ordinary differential system of $7 \times M$ equations, in terms of the coordinate $\zeta$. Among them, $4 \times M$ are ordinary differential equations for the spectral coefficients $u_{\ell}$, $\pi'_{\ell}$, $\Phi'_{\ell}$ and $d\Phi'_{\ell}$ ({\it i.e.} Eqs. \ref{App_eq_Mouvement}, \ref{App_eq_Continuite}, \ref{App_eq-aux}, and \ref{App_eq_Poisson2}), and the remaining $3 \times M$ equations are algebraic equations for the spectral coefficients $v_{\ell}$,  $w_{\ell_p}$ and  $\rho'_{\ell}$ ({\it i.e.} Eqs. \ref{App_eq_Divergence}, \ref{App_rot}, and \ref{App_eq_relat_adiab2}).

We choose to solve this system by an Newton-like method, which consists in taking a guess at the pulsation frequency $\sigma_0$ and looking for the pulsation mode with the closest frequency to this guess. Solving the system yields a deviation, $\delta \sigma$, from the initial guess, $\sigma=\sigma_0+\delta \sigma$ is taken as a new guess for the next iteration, and this process is iterated till convergence (three steps are in general enough).

We therefore isolate the terms proportional to $\delta \sigma$, and obtain the following system of equations which can be put under the form of a matrix:

\begin{align}
\frac{ dy_1}{ d \zeta} \, &= \, ( A_{11} \, + \, \delta \sigma \,  A_{12}) \,  y_1 \, + \, ( A_{21} \, + \, \delta \sigma  A_{22}) \,  z_1 \, , \\
0 \, &= \, ( B_{11}\, + \, \delta \sigma \,  B_{12}) \,  y_1 \, + \, ( B_{21} \, + \, \delta \sigma \,  B_{22}) \,  z_1\, , 
\label{substit}
\end{align}
where $y_1$ and $z_1$ are column vectors containing the spectral coefficients of the unknowns
\begin{equation}
 y_1\, =\, \begin{bmatrix} \pi'_{\ell} \vspace{0.1cm} \\ \Phi'_{\ell} \vspace{0.1cm} \\ \rm d\Phi'_{\ell} \vspace{0.1cm} \\ u_{\ell} \end{bmatrix}  \hspace{1cm} \hbox{and} \hspace{1cm}  z_1= \begin{bmatrix} v_{\ell} \vspace{0.1cm} \\ w_{\ell_p}  \vspace{0.1cm} \\ \rho'_{\ell} \end{bmatrix}\, , 
\label{vect_ppe}
\end{equation}
and where $\ell \, = \, \mid m \mid \, + \, 2(j-1) \, + \, par, \,  \forall j \in [ 1, M ], $.\\
$A_{11}$, $A_{12}$, $A_{21}$ and $A_{22}$ correspond to the following equations
\begin{equation}
\begin{bmatrix} \hbox{ \, pseudo-radial motion (Eq. \ref{App_eq_Mouvement}) \, } \\ \hbox{ \, definition of $d\Phi'$ (Eq. \ref{App_eq-aux})  \, } \\ \hbox{ \, Poisson (Eq. \ref{App_eq_Poisson2}) \, } \\ \hbox{ \, continuity (Eq. \ref{App_eq_Continuite}) \, } \end{bmatrix} ,
\end{equation}
whereas $B_{11}$, $B_{12}$, $B_{21}$ and $B_{12}$ correspond to the algebraic equations
\begin{equation}
\begin{bmatrix} \hbox{ \, poloidal motion (Eq. \ref{App_eq_Divergence}) \, }\\ \hbox{ \, toroidal motion (Eq. \ref{App_rot}) \, }\\ \hbox{ \, adiabatic relation (Eq. \ref{App_eq_relat_adiab2}) \, } \end{bmatrix} .
\end{equation}
Thanks to the three last equations, the non differentiated unknowns can be expressed in terms of the differentiated ones with the help of Eq. (\ref{substit}). To do so, the matrix $(B_{21} \, + \, \delta \sigma \,  B_{22})$, which is a real square matrix of rank $3M$, has to be inverted.
It is straightforward to show that if $\delta \sigma$ is small enough, 
\begin{equation}
\hspace{-0.25cm}\left( \rm B_{21} \, + \, \delta \sigma \, \rm B_{22}\right)^{-1} \, = \, \rm B_{21}^{-1} \, - \, \delta \sigma \, \rm B_{21}^{-1} \, \rm B_{22} \, \rm B_{21}^{-1} \, + \, o(\delta \sigma^2) \, .
\end{equation}
Thus, the matrix system can be written under the form: 
\begin{align}
\label{syst_mat}
\frac{dy_1}{d\zeta} \, = \, \left(A \, + \, \delta \sigma \, A_{\delta \sigma} \right) \, y_1\\
z_1 \, = \, ( B \, + \, \delta \sigma \, B_{\delta \sigma} ) \, y_1
\end{align}

\subsection{Radial discretization}
\label{ss_discret}
In the radial direction, structural quantities, and as a consequence eigenmodes, undergo sharp variations (for instance at the top of a convective core, or at the star's surface). Therefore, in the radial direction, spectral methods are inappropriate when dealing with realistic stellar models. We choose to discretize the differential equations using a finite differences approach. The continuous domain of integration is replaced by a discrete set of $N_r$ radial points. The number of points in the radial grid is an important factor which affects the numerical precision, as discussed in detail in Sect. \ref{ss_conv}. Another characteristic of the discretization scheme, which comes into play in the numerical precision is the estimation of the derivatives. With classical finite differences methods, the more precise you get the less stable are the computations.

We have adopted a fourth-order difference scheme, which relies on the following identity
  
\begin{equation}
\label{diff_scuf}
y_i+\frac{h}{2} y'_i+ \frac{h^2}{12} y''_i = y_{i+1}-\frac{h}{2} y'_{i+1}+\frac{h^2}{12} y''_{i+1} + o\left(h^5\right) \,,
\end{equation}
where the primes denote derivatives with respect to $\zeta$, $h$ is given by $h=\zeta_{i+1}-\zeta_i$, and the subscripts ``$_i$'' and ``$_{i+1}$'' denote the layer indexes.
The great advantage of this scheme is that it is accurate up to $h^4$, while retaining high numerical stability, as it only involves two consecutive grid points. This finite differences scheme has already been implemented by \cite{Scuflaire2008} in the Li\`ege Oscillation Code which has been proven to be very stable and accurate for the calculations of all types of pulsation modes in all kinds of stars.

From now on, we drop the subscripts ``$_1$'' from $y_1$. Thanks to Eq. (\ref{syst_mat}), it is possible to express the derivatives of $y$. The identity (\ref{diff_scuf}) can then be valid as long as the matrices coefficients in $A$ and $A_{\delta \sigma}$ are continuous and have continuous derivatives. We then get the following matrix equation at each layer $i$
\begin{equation}
  \alpha^+_i  y_i \, - \, \alpha^-_{i+1} y_{i+1} \, = \, \delta \sigma \, \left[ \beta^+_i y_i \, + \, \beta^-_{i+1} y_{i+1}\right]\, + \, o\left(h^5 \right)\, ,
\end{equation}
where
\begin{align}
\alpha^+_{i} \, &= \, I_d \, + \, \frac{h}{2}\, A_i \, + \, \frac{h^2}{12} \, \left(A^2_i \, + \, A'_i\right)\, , \nonumber \\
\beta^+_{i} \, &= \, \frac{h}{2} \, A_{\delta \sigma, \, i} \, + \, \frac{h^2}{12} \, \left(A_i \, A_{\delta \sigma, \, i} \, + \, A_{\delta \sigma, \, i} \, A_i \, + \, A'_{\delta \sigma, \, i} \right)\, , \nonumber \\
\alpha^-_{i} \, &= \, I_d \, - \, \frac{h}{2} \, A_{i+1} \, + \, \frac{h^2}{12} \, \left(A^2_{i+1} \, + \, A'_{i+1}\right)\, , \nonumber \\
\beta^-_{i} \, &= \, - \, \frac{h}{2} \, A_{\delta \sigma, \, i+1}  \nonumber \\
&+ \, \frac{h^2}{12} \, \left(A_{i+1} \, A_{\delta \sigma, \, i+1} \, + \, A_{\delta \sigma, \, i+1} \, A_{i+1} \, + \, A'_{\delta \sigma, \, i+1} \right)\, , \nonumber
\end{align}
Here, $I_d$ is the square  $4 M \times 4 M$ identity matrix.
The system of equations over the whole stellar interior can then be put under the canonical form:
\begin{equation}
\mathcal{A} \, \mathcal{Y} \, = \, \delta \sigma  \, \mathcal{A}_{\delta \sigma} \, \mathcal{Y} \, ,
\label{pb_lin}
\end{equation}
 where the vector $\mathcal{Y}$ has been introduced:
\begin{equation}
\mathcal{Y} \, = \, \begin{bmatrix}  y_1 \\  y_2 \\ \vdots \\ y_N  \end{bmatrix} \hspace{1cm} (\hbox{layers } 1,\, \cdots N_r)
\end{equation}
$\mathcal{A}$ and $\mathcal{A}_{\delta \sigma}$ being block diagonal matrices.

We impose boundary conditions which can be expressed as algebraic relations, as explained in detail in Appendix (\ref{app_cond-cent}).  

We note that Eq. (\ref{pb_lin}) is a generalization of the classical eigenvalue problem $A y = \lambda y$.

\subsection{Inverse iteration algorithm}

In order to calculate the eigenvectors, $\mathcal{Y}$, and eigenvalues, $\delta \sigma$, we generalize the classical inverse iteration algorithm to the more general problem formulated in Eq. (\ref{pb_lin}), as developed by \cite{Dupret2001} in the stellar pulsations context.

The estimate at step $k+1$ of the eigenvector, $\mathcal{Y}_{k+1}$, is obtained from the estimate at step $k$ and the initial guess, $\delta \sigma_0$, of the eigenvalue by the formula
\begin{equation}
\mathcal{Y}_{k+1} \, = \, ( \mathcal{A} \, - \,\delta \sigma_0 \, \mathcal{A}_{\delta \sigma} )^{-1} \, \mathcal{A}_{\delta \sigma} \, \mathcal{Y}_{k} \,.
\end{equation}
If we assume that $\mathcal{A}$ is inversible and that $\mathcal{A}^{-1} \, \mathcal{A}_{\delta \sigma}$ is diagonalizable, it is easy to prove that this algorithm has the same geometrical convergence rate as the classical inverse iteration algorithm. The inverse matrix is not explicitly calculated but we solve the system:
\begin{equation}
( \mathcal{A} \, - \,\delta \sigma_0 \, \mathcal{A}_{\delta \sigma} ) \, \mathcal{Y}_{k+1} \, = \,  \mathcal{A}_{\delta \sigma} \, \mathcal{Y}_{k} .
\end{equation}
 To do so, we perform a LU factorization of the system with partial pivoting. Then, we iterate solving the 2 triangular systems at each step. 
Note that if the initial guess for the eigenvalue is good, the algorithm converges quickly towards the solution even with a bad eigenvector as an initial estimate.

The corresponding eigenvalue is then determined by the generalization of the Rayleigh ratio
\begin{align}
\label{rap_ray}
\delta \sigma \, = \, \frac{\mathcal{Y}^* \, \mathcal{A}_{\delta \sigma}^* \, \mathcal{A} \, \mathcal{Y}}{\mathcal{Y}^* \, \mathcal{A}_{\delta \sigma}^* \, \mathcal{A}_{\delta \sigma} \, \mathcal{Y}} \, ,
\end{align}
where $\mathcal{Y}^* $ and $ \mathcal{A}_{\delta \sigma}^*$ are the Hermitian conjugates of $\mathcal{Y}$ and $ \mathcal{A}_{\delta \sigma}$, respectively. It can be easily shown that $\delta \sigma$, given by Eq. (\ref{rap_ray}), minimizes:
\begin{align}
\label{mini_rap_ray}
S^2\, &= \, \mathcal{Y}^* \,(\mathcal{A}^* \, - \, \delta \sigma \, \mathcal{A}_{\delta \sigma}^*) \, (\mathcal{A} \, - \, \delta \sigma \, \mathcal{A}_{\delta \sigma}) \, \mathcal{Y}\nonumber \\
&= \, \mid \mid \left( \mathcal{A} \, - \, \delta \sigma \, \mathcal{A}_{\delta \sigma}\right) \, \mathcal{Y}\mid \mid 
\end{align}

\section{Tests and accuracy}
\label{S4}

As mentioned in the introduction, in order to be integrated to a stellar modeling chain for massive computations, the program has been developed with a constant concern for simplicity and rapidity of use. In this section, we assess the role of numerical parameters in the convergence process toward an oscillation mode, and establish the computational performances with respect to these parameters. 

\noindent What takes up the most computing resources in ACOR is the integrations over $\theta$ of the equation coefficients (Eq. \ref{eq_principe_decomp}). These coefficients are evaluated at each radial layers (for $N_r$ layers) by projecting the equations onto the spherical harmonics basis (of dimension $M$). Therefore, by assessing the role of the two parameters $M$ and $N_r$, it is possible to define the optimal values for a good compromise between computation time, memory resources and accuracy of the results.

Note that there is no automatic method that allows us to identify the modes. In this work, we followed the progression of the modes, as we gradually increased the rotation rate from zero to a high value.  We used the kinetic energy distribution in the meridional plane, such as in Fig. \ref{avcr_diag2D}, to correctly select the mode at each step.

All the tests presented in this paper are made assuming uniformly rotating polytropes of polytropic index $N=3$ (polytropic exponent $\gamma=4/3$) with an adiabatic index of $\Gamma_1 \, = \, 5 / 3$.

\subsection{Convergence tests}
\label{ss_conv}

\begin{figure} [t!]
\begin{center}
\hspace*{-1cm}\includegraphics[scale=0.28, angle=-90]{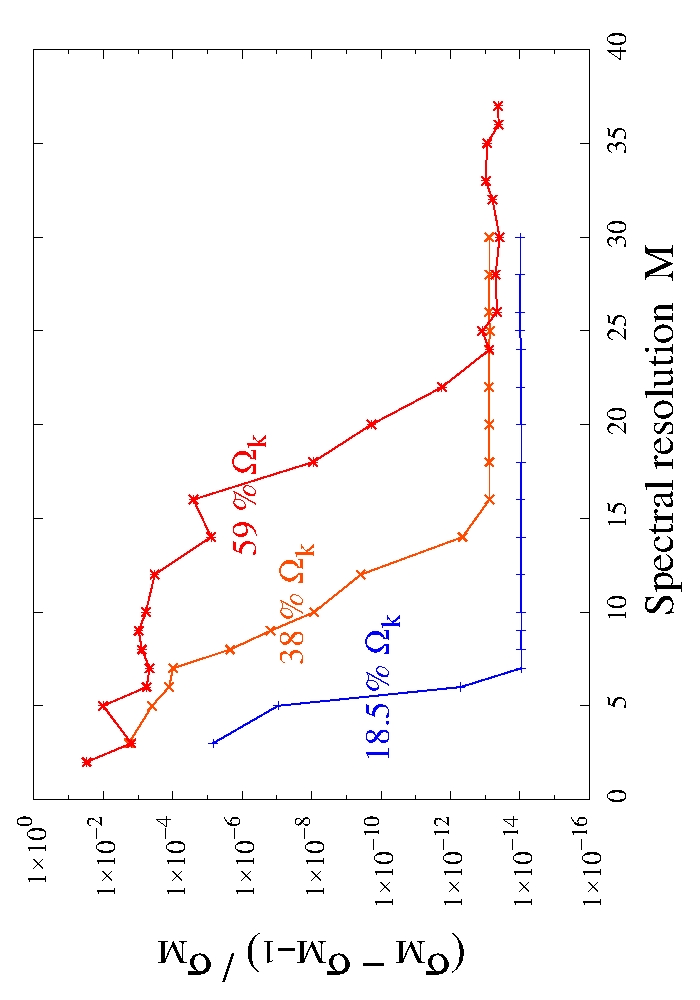}
\caption{\label{Conv_M}Convergence as a function of the number of spherical harmonics, taken as the relative frequency difference between computations using consecutive spectral resolutions, M-1 and M, for an $n=3$ mode dominated by an $\ell=1$, $m=0$ component, and for three different rotational velocities: $18.5 \% \, \Omega_k$ in blue, $38 \% \, \Omega_k$ in orange and $59 \% \, \Omega_k$ in red.}
\end{center}
\end{figure}

In the ideal case where equilibrium quantities would be $\mathcal{C}^{\infty}$,  the numerical errors would decrease as $e^{-aM}$. Rotation induces non-spherical profiles for the equilibrium quantities, and causes the eigenfunctions to depart from a single spherical harmonic. Therefore, the higher the rotation rate, the stronger the deviations from sphericity, and the larger the spherical harmonic basis has to be, as illustrated indeed, in Fig. \ref{Conv_M}. The convergence calculations illustrated in Fig. \ref{Conv_M} show that convergence is reached for 7 terms for $\Omega=18.5 \%\Omega_k$, 16 terms for  $\Omega=37.9 \%\Omega_k$ and 25 terms for $\Omega=58.9 \%\Omega_k$ in the spectral expansion. This figure also shows that the convergence reaches machine precision.

\begin{figure} [t!]
\begin{center}
\hspace*{-1cm}\includegraphics[scale=0.28, angle=-90]{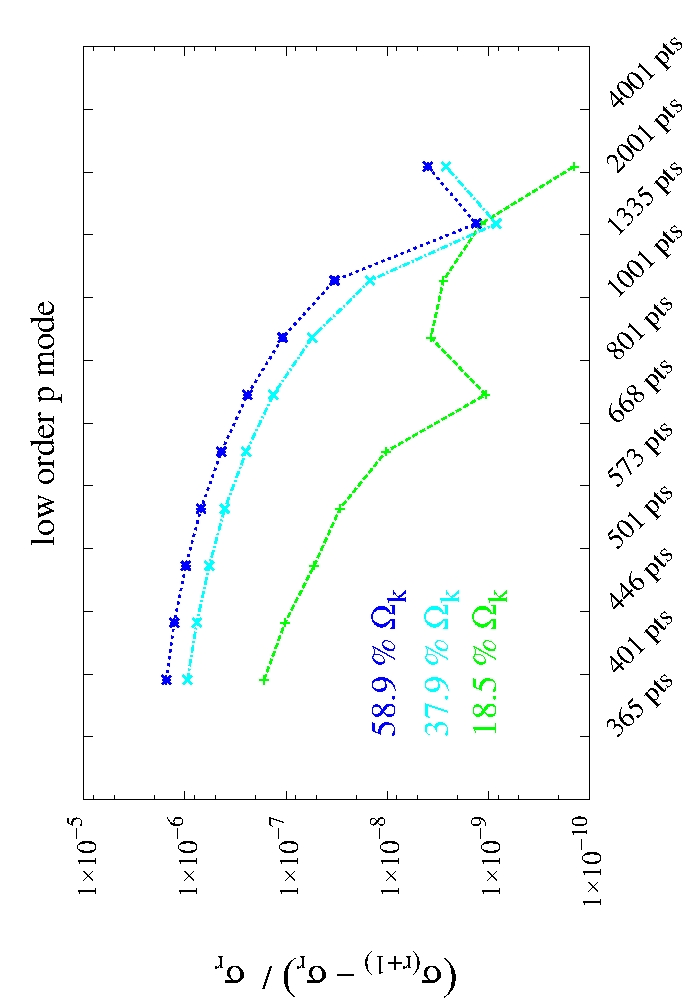}
\hspace*{-1cm}\includegraphics[scale=0.28, angle=-90]{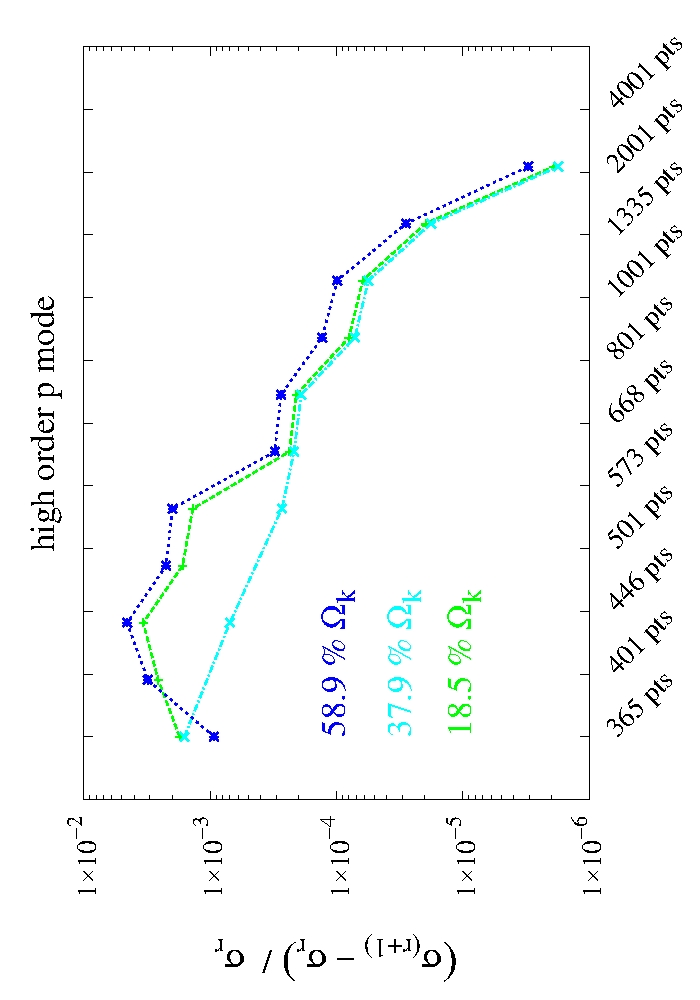}
\caption{\label{Conv_R} Convergence as a function of the radial resolution, taken as the relative frequency differences between computations using consecutive radial resolutions, r and (r+1) points, at three different rotational velocities: $18.5 \% \, \Omega_k$ in green, $37.9 \% \, \Omega_k$ in light blue and $58.9 \% \, \Omega_k$ in dark blue. In both panels, the mode is dominated by the $\ell=2$, $m=0$ component. The upper panel corresponds to an $n=3$ mode and uses irregular grids, whereas the lower panel shows an $n=9$ mode calculated with even distributions of points.}
\end{center}
\end{figure}

Concerning the convergence with respect of the radial resolution, we present in Fig. \ref{Conv_R} the worst (bottom) and the best case (top). The higher the radial order $n$, the more numerous the nodes are in the eigenfunction and the higher the radial resolution has to be. These plots also show that using a non-regular radial grid, whose number of points increases as we go outwards, allows us to increase the accuracy of computations. This is due to the fact that the computed modes are acoustic modes, with consequently a small wavelength in the outer layers.

\begin{table}[t!]
\caption{\label{tab_perf} Numerical resources -- $\mathrm{i.e.}$ time in minutes and memory in MB or GB -- used by ACOR with a spectral resolution M and a radial one of N$_r$.
}
\vspace{0.5cm}
\centering
\begin{tabular}{lccccc}
\hline
 N$_r$  &  M  & Matrix size & Time & memory \\
\hline
\hline
 2500 &  40 & $96 \times 10^{6}$ & 17 min 20 s & 6.4 GB \\
 2500 &  20 & $24 \times 10^{6}$ & 2 min 30 s & 1.6 GB \\
 2500 &  10 & $6 \times 10^{6}$ & 24 s  & 500 MB \\
\hline
 1250 &  40 & $48 \times 10^{6}$ & 8 min & 3.2 GB \\
 1250 &  20 & $12 \times 10^{6}$ & 1 min 10 & 850 MB \\
 1250 &  10 & $3 \times 10^{6}$ &  12 s & 250 MB \\
\hline
 625 &  40 & $24 \times 10^{6}$ & 3 min 50 & 1.6 GB \\
 625 &  20 & $6 \times 10^{6}$ & 33 s & 425 MB \\
 625 &  10 & $1.5 \times 10^{6}$ & 6 s & 125 MB \\
\hline
\end{tabular}
\end{table}

\subsection{Numerical resources}

Among all the operations performed by the code, calculating the projection integrals (Eq. \ref{eq_principe_decomp}) is the most demanding in terms of computational time, whereas inverting the two matrices $ \mathcal{A}$  and $\mathcal{A}_{\delta \sigma}$ requires the most memory. In Table \ref{tab_perf} are indicated the memory and time resources needed by the computations with respect to the parameters $M$ and $N_r$. Note that $ \mathcal{A}$  and $\mathcal{A}_{\delta \sigma}$ are bloc matrices, their size corresponds to the number of non-zero elements they contain.

\section{Comparison with Reese et al. (2006)}
\label{S5}
\begin{figure}[t!]
\begin{center}
\includegraphics[scale=0.28, angle=-90]{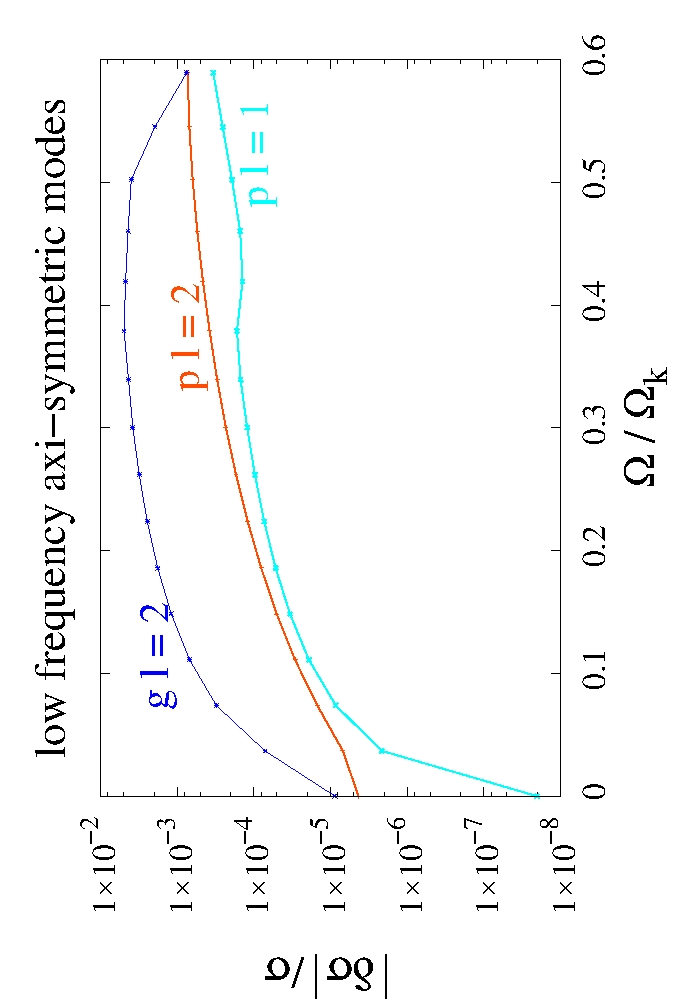}
\includegraphics[scale=0.28, angle=-90]{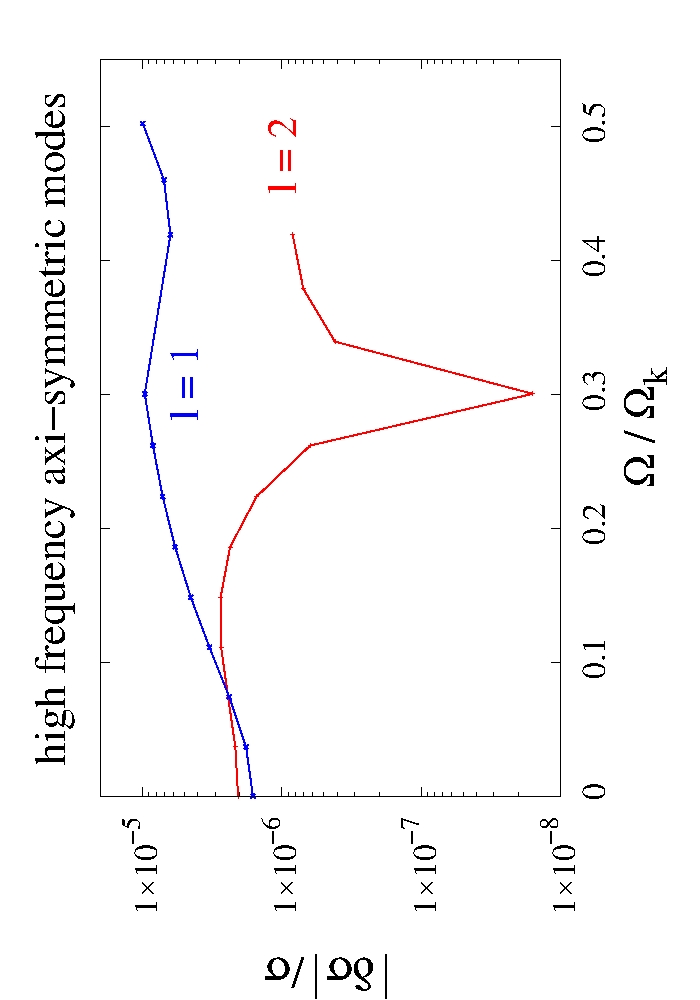}
\caption{\label{fig_diff_freq} Discrepancies between eigenfrequencies computed by ACOR and TOP as a function of the rotation angular velocity for five axisymmetric modes: \textit{Top:} for two $n=3$ modes, corresponding to $\ell=1$ (light blue) and $\ell=2$ (orange) in the non rotating case and one $n=1$ g mode with $\ell=2$; \textit{Bottom:} $n=9$ modes, corresponding to $\ell=1$ (dark blue) and $\ell=2$ (red).}
\end{center}
\end{figure}

\begin{figure}[t!]
\begin{center}
\includegraphics[scale=0.33, angle=-90]{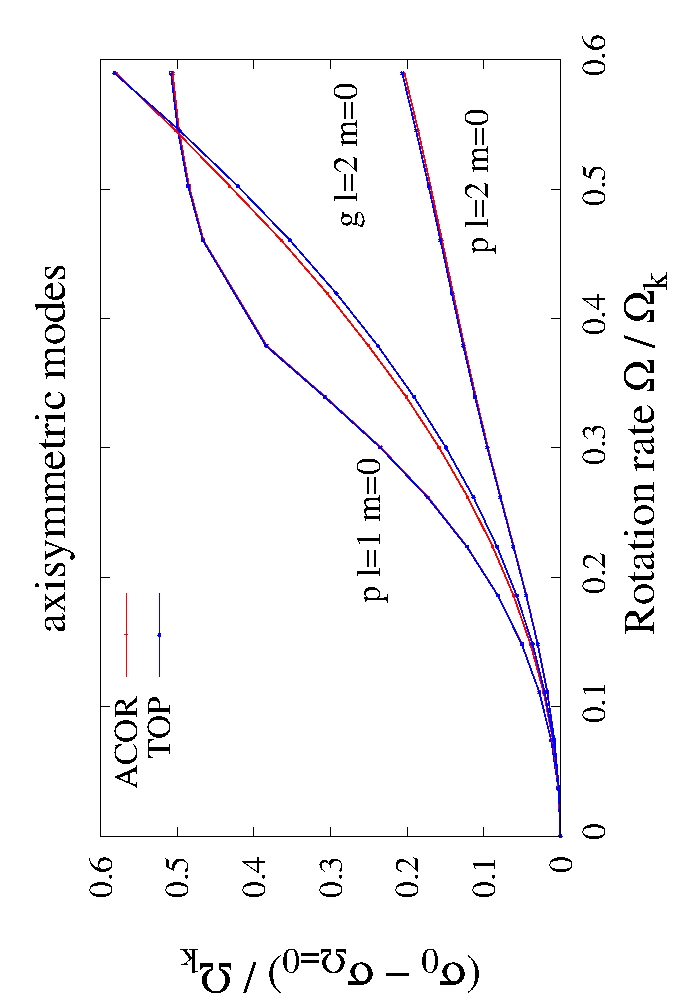}
\includegraphics[scale=0.33, angle=-90]{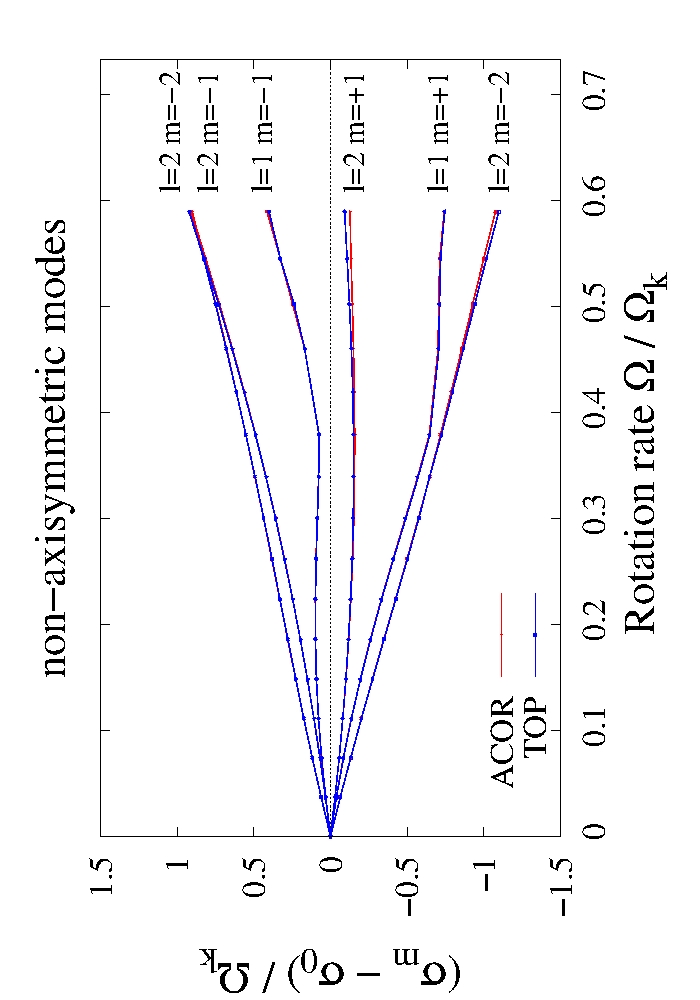}
\caption{\label{fig_comp} Behavior of the eigenfrequencies computed by ACOR and TOP with respect of the rotational angular velocity for two multiplets; \textit{Top:} centroid modes with dominant component $\ell=1, \, m=0$, and $\ell=2, \, m=0$, \textit{Bottom:} sectorial and tesseral modes, dominated by components $\ell=1, \, \rm m=\pm1$ and $\ell=2, \, \rm m=\pm1$ and $\pm2$.}
\end{center}
\end{figure}
After the numerical tests presented in Sect. \ref{S4}, the aim of this section is to validate ACOR's results by comparing them with those from the TOP code (Two-dimensional Oscillation Program). The TOP code has been developed by \cite{Reese2006} for two dimensional polytropes as a first step. The approach is based on a two-dimensional spectral method which uses Tchebichev polynomials for the radial dependence. We present here the comparison between TOP and ACOR results for identical polytropic models.

Roughly, a polytrope is a simplified model for which one assumes an ad-hoc relation between density ($\rho_0$) and pressure ($p_0$):
\begin{align}
p_0\, = \, K \, \rho_0^{\gamma}\, ,
\end{align}
where $K$ is the polytropic constant and $\gamma$ is called the polytropic exponent which has been taken as $4 / 3$ here. The detailed method used to compute the rotating polytropic models is given in \cite{Rieutord2005}. This method is an iterative scheme based on a spectral decomposition using Tchebichev polynomials for the radial dependence, and spherical harmonics for the horizontal one. We subsequently interpolate these models onto a radial grid which is appropriate for finite differences.

Concerning the parameters used for the calculations within the two codes, the angular spectral resolutions have been fixed so as to reach convergence. It therefore depends on the rotation velocity and varies from 10 terms in the harmonics expansion for low rotation rates to 25 for the highest ones. For TOP, the spectral resolution in terms of the Tchebichev polynomials varies from 50 to 80 terms. For ACOR, the radial resolution is fixed at $N_r = $2000 grid points.

\subsection{Eigenfunctions comparison}

The major effect of centrifugal distortion is the loss of spherical symmetry, which results in the coupling of spherical harmonics of different degrees to describe the horizontal behavior of a single mode. In Appendix \ref{App_comp_modes} are given the contributions of dominant spherical harmonics in the spectral expansion of two modes: an odd mode dominated by an $\ell=1$, $m=0$ component (see Fig. \ref{fig_Pi-l_10_bf2}) and an even one dominated by $\ell=2$, $m=0$ (see Fig. \ref{fig_Pi-l_20_bf2}). The solid lines correspond to the calculations done with ACOR and the dotted lines to those done with TOP. These plots clearly show that the evolution of the angular behavior of the modes with respect to rotation obtained by the two codes is very similar. This allows us to validate the analytical calculations as well as the inverse iteration algorithm, which converges onto eigenfunctions, while eigenfrequencies are computed a posteriori through the minimization of Eq. (\ref{mini_rap_ray}). 

\subsection{Eigenfrequencies comparison}

Concerning the comparisons of eigenfrequencies, Fig. \ref{fig_diff_freq} shows the frequency differences between the two codes, for odd and even eigenmodes in the low (top) and high (bottom) frequency regimes.\textcolor{black}{} Globally, the discrepancies between results from the two codes is of the order of $0.0001 \% - 0.08 \%$ ($3 \times 10^{-3} \, \Omega_k$ at most) for p modes, and   $0.5\%$ for g modes ($1 \times 10^{-2} \, \Omega_k$ at most) which seems very satisfying considering that the two codes rely on different and independent computations, and in particular a different treatment for the central boundary conditions. Once more, this makes numerical programming mistakes unlikely in our calculations.

One of the observational characteristics of rotation in seismology is the rotational generalized splitting, $\mathrm{i.e.}$ the frequency difference between two modes with the same radial order and harmonic degree, but with opposite azimuthal orders. Figure \ref{fig_comp} shows the evolution of the structure of two multiplets with respect to the rotation velocity.  The plots in the bottom panel show that the two codes find the same structure for the multiplets regardless of the rotation rate (from 0 to $60 \% \, \Omega_k$) and the symmetry class of modes. The impact of rotation on centroid modes (top panel) is also the same with the two methods.

The agreement between the two oscillation programs developed separately, not only on the eigenfunctions and on frequencies, but also on the structure of the spectra, allows us to validate the approach adopted by ACOR.

\section{Illustration: avoided crossing}
\label{S6}
\begin{figure} [t!]
\begin{center}
\hspace*{-1cm}\includegraphics[scale=0.6, angle=0]{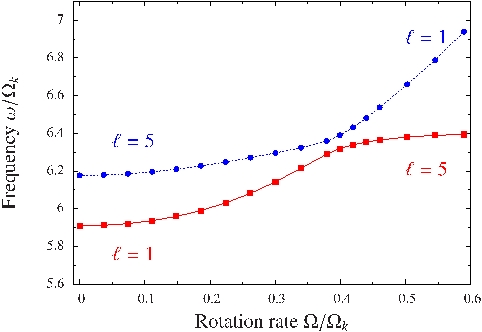}
\caption{\label{avcr_15}Avoided crossing, illustrated by plotting the frequency with respect to the rotation rate for two $n=2$ modes (referred to as mode $\#1$ in red and mode $\#2$ in blue), which, in the non-rotating case, are $\ell=1$ and $\ell =5$. The harmonic degrees given in the graph is the degree of the dominant component for each mode.}
\end{center}
\end{figure}In quantum mechanics, an avoided crossing (or level repulsion) occurs, for instance, in a two level system ($|1\rangle$ and $|2\rangle$), placed in a magnetic field which acts differently on the two levels \citep[see for example ][]{Cohen-Tannoudji1973}. When the two states are coupled, the levels repulse each other, since the system's energy cannot be degenerated.

Accordingly, in a rotating star, pulsation mode frequencies tend to cross because the modes are not affected the same way by rotation \citep[particularly by the centrifugal force, see ][]{Lignieres2006}. As two modes of the same parity cannot have the same frequency, an avoided crossing occurs during which the two modes exchange their characteristics.

\begin{figure*} [p]
\begin{center}
\includegraphics[scale=1, angle=180]{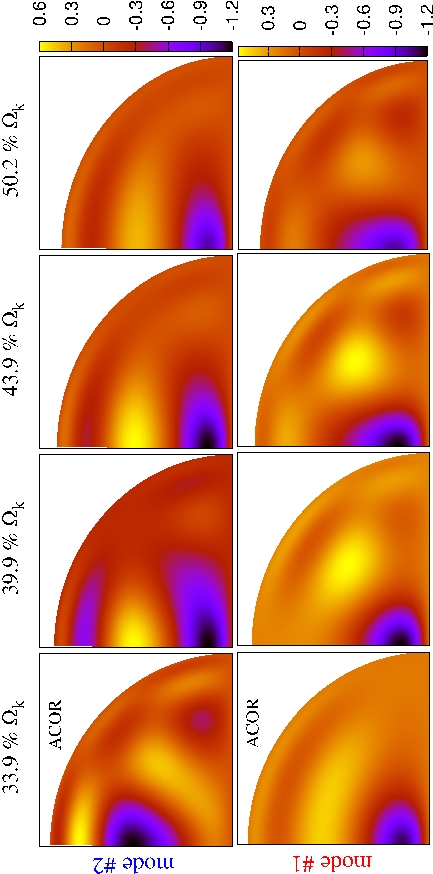}
\caption{\label{avcr_diag2D}Spatial distribution of the pressure perturbation in a meridional plane for two low frequency modes (radial order $n=2$, azimuthal order $m=0$) involved in an avoided crossing occurring around $\Omega = 40 \% \Omega_k$, as modeled by ACOR.  \textit{Left:} at low rotation velocities, the mode $\#1$ is dominated by an $\ell=1$ component and changes nature to become dominated by $\ell=5$ as the velocity increases.  \textit{Right:} at low rotation velocities, the mode $\#2$ is dominated by an $\ell=5$ component and changes nature to become dominated by $\ell=1$ as the velocity increases.}
\end{center}
\end{figure*}

This is illustrated in Fig. \ref{avcr_15} by the evolution of the frequencies of two coupled modes with respect to the rotation rate. The two modes are of the same symmetry with $m=0$, and $\ell=1$ or 5. As explained earlier, their frequencies cannot be degenerated, therefore the crossing is avoided, and as shown in Fig. \ref{avcr_diag2D} they progressively exchange angular characteristics. With the help of the distribution of the pressure perturbation in the meridional plane, we show that the mode with geometry dominated by $\ell=1$ at moderate rotation rates (mode $\#1$), ends up with a dominant $\ell=5$ component at high rotation rates. 

Therefore, this work \citep[see also][]{Reese2009} shows that modes in rapidly rotating stars can no longer be identified only by one angular degree $\ell$. Actually, when rotation increases, the different components in the spectral expansion are more and more coupled by the non-spherical terms of the system of equations. As a consequence, each mode is composed of a mixture of spherical harmonics of the same symmetry, and it is not even possible to follow a mode considering its dominant component, as it can change during an avoided crossing.  

\section{Conclusion}
\label{S7}
A new oscillation code, which computes non-radial adiabatic pulsations in rotating stars has been developed. The accuracy of the calculations has been achieved thanks to a hybrid method based on a spectral expansion on the spherical harmonics basis and a fourth-order finite differences scheme. The code has been tested and validated in the present study for polytropic models, but no assumptions were made in the implementation concerning the structure of the model used as input to ACOR. 

Although we've limited ourselves to barotropic models in this paper, it must be emphasized that our code is fully able to handle non-barotropic rapidly rotating models as will be presented in a forthcoming paper.
This is an important point for studying the pulsations of realistic stellar models such as those including rotationally induced transport processes according to \cite{Zahn1992}.  Indeed, such processes lead to shellular rotation profiles in radiative zones, which are as a consequence non barotropic.
Moreover the radial treatment based on a finite differences method which is accurate up to the fourth order is particularly well suited for stellar models presenting sharp variations of the structural quantities.

\begin{acknowledgements}
R-M.O. is indebted to the ``F\'ed\'eration Wallonie-Bruxelles - Fonds Sp\'eciaux pour la Recherche / Cr\'edit de d\'emarrage - Universit\'e de Li\`ege'' for financial support. D.R.R. is financially supported through a postdoctoral fellowship from the ``Subside f\'ed\'eral pour la recherche 2011'', University of Li\`ege. The authors would like to thank J. Ballot for providing g modes frequencies.
\end{acknowledgements}

\begin{onecolumn}
\begin{appendix}

\section{Equation system}

Rather than projecting the motion equation onto the basis vectors, we choose to decompose it into one radial, one poloidal and one toroidal field. We obtain one equation without derivatives relative to the angular coordinates by projecting the motion equation onto $\overrightarrow{e_{r}}$. Then by taking the divergence, and the pseudo-radial componant of the curl of the motion equation, we get two equations without radial derivatives. 

\subsection{Radial motion}
\begin{align}
\label{App_eq_Mouvement}
\frac{\partial \pi'}{\partial \zeta}\, =  \, &-\,  i \, \left( m \, \Omega \, + \, \sigma \right) \, M_{\zeta} \, v'_{\zeta} \, -\, i \, \left( m \, \Omega \, + \, \sigma \right) \, \sin \theta \, M_{\mu} \, \tilde{v}'_{\theta} \\
&- \,\sin\theta \,M_{\phi} \, v'_{\phi} \,- \,\frac{\partial \Phi'}{\partial \zeta} \,+ \,M_{\rho'} \, \rho' \, + \,M_{\pi'} \,\pi' \, ,\nonumber\\
\hbox{with,  }&&  \, ,\nonumber \\
&M_{\zeta}= r_{\zeta} g(\zeta,\mu)   \, ,\nonumber \\
&M_{\mu}=-r_{\zeta} g(\zeta,\mu) \frac{r_{\mu}}{r}   \, ,\nonumber \\
&M_{\phi}=-2\Omega r_{\zeta}  \, ,\nonumber \\
&M_{\rho'}=\frac{1}{\rho^2} \left. {\frac{\partial p}{\partial \zeta}} \right)_{\theta,\varphi}  \, ,\nonumber \\
\label{App_eq_M_mu}
&M_{\pi'}= -\frac{1}{\rho} \left. {\frac{\partial \rho}{\partial \zeta}} \right)_{\theta,\varphi} \, ,
\end{align}
where the colatitude $\theta$ is replaced by  $\mu \, = \, \cos \theta$, so $r_{\theta} \, = \, - \, \sin \theta \, r_{\mu}$. As stated in Sect.~\ref{ss_basic-eq}, we applied the change of variable $\pi' \, = \, p' \, / \, \rho$ in order to avoid singularity problems at the surface of polytropic models.

\subsection{Poloidal motion}

Here we calculate the divergence of the horizontal component of the motion equation. To do so, one has to apply the operator $r \vec{\nabla}_{\perp} \cdot$, where for any vector $\vec{X}=X_{\zeta}\overrightarrow{a_{\zeta}}+X_{\theta}\overrightarrow{a_{\theta}}+X_{\varphi}\overrightarrow{e_{\varphi}}$:
\begin{align}
\vec{\nabla}_{\perp} \cdot \vec{X}= \, \frac{1}{r \sin \theta} \, \frac{\partial }{\partial \theta} \left( \sin \theta X_{\theta}\right)_{\zeta} \, + \, \frac{1}{r \sin \theta} \, \frac{\partial }{\partial \varphi} \left( X_{\varphi}\right) \, . 
\end{align}
Once applied to the motion equation, this gives
\begin{align}
&\frac{i}{\sin \theta} \, \frac{\partial}{\partial \theta} \, \left( \sin \theta \, (m \Omega + \sigma) v'_{\theta} \right)_{\zeta} \, - \, \frac{2}{\sin \theta} \, \frac{\partial}{\partial \theta} \, \left( \Omega \sin \theta \, (\frac{r_{\theta}}{r} \, \sin \theta \, + \, \cos \theta) v'_{\phi} \right)_{\zeta} \, - \, \frac{m \,(m \Omega + \sigma)}{\sin \theta} v'_{\phi}  \nonumber \\
&+ \, 2 i m \Omega g \, \left( \cot \theta \, + \, \frac{r_{\theta}}{r} \right) \, v'_{\theta} \, + \, i m g \, \left[ 2 \Omega \,  \left( 1  - \frac{r_{\theta}}{r} \, \cot \theta \right) \,+ \, \frac{r}{r_{\zeta}} (1 - \frac{ r_{\theta}^2}{ r^2})\, \frac{\partial \Omega}{\partial \zeta} \, - \, \frac{ r_{\theta}}{ r} \, \frac{\partial \Omega}{\partial \theta}\right] \, v'_{\zeta}\, + \,  i m  g \, \frac{\partial \Omega}{\partial \theta} v'_{\theta}   \nonumber \\
&= \, \frac{1}{ r} \, L^{2} \pi' \, + \, \frac{1}{r} \, L^{2} \Phi' \, + \, \frac{ r_{\theta}}{ r^{2}} \frac{\partial \pi'}{\partial \theta}\, + \, \frac{ r_{\theta}}{ r^{2}} \frac{\partial \Phi'}{\partial \theta} \, - \, \frac{1}{\sin \theta} \frac{\partial }{\partial \theta} \, \left( \frac{\sin \theta}{ r \rho} \,\frac{\partial \rho}{\partial \theta} \, \pi' \right)_{\zeta} \, + \, \frac{1}{\sin \theta} \frac{\partial }{\partial \theta} \, \left( \frac{\sin \theta}{ r \rho^2} \,\frac{\partial  p}{\partial \theta} \, \rho' \right)_{\zeta} \, ,
\end{align}
where $g$ is given by $g(\zeta,\mu) = 1 / (1+(1-\mu^2)r^2_{\mu}/r^2) $,  and the operator $L^2$ is the angular momentum operator:
\begin{align}
L^2 \Psi \, &= \, - \, \frac{1}{\sin \theta} \, \frac{\partial }{\partial \theta} \, \left. { \sin \theta  \, \frac{\partial \Psi }{\partial \theta}} \right)_{\zeta} \, - \, \frac{1}{\sin^2\theta} \, \frac{\partial^2 \Psi}{\partial \varphi^2} \nonumber \\
&= \, - \, (1-\mu^2) \, \left. { \frac{\partial^2 \Psi}{\partial \mu^2}} \right)_{\zeta} \, + \, 2 \mu  \left. { \frac{\partial \Psi}{\partial \mu}} \right)_{\zeta} \, - \, \frac{1}{(1-\mu^2)} \frac{\partial^2 \Psi}{\partial \varphi^2}\, ,
\end{align}
whose eigenfunctions are the spherical harmonics: $L^2 \, Y_{\ell}^{m}(\theta,\varphi) \, = \, \ell \, (\ell+1) \,  Y_{\ell}^{m}(\theta,\phi)$.
Thus, the equation becomes
\begin{align}
\label{App_eq_Divergence}
0\, &= \, (m \Omega + \sigma) \, \left[ \frac{ i}{\sin \theta} \, \frac{\partial}{\partial \theta} \, \left( \sin \theta \,  v'_{\theta}  \right) \, - \, \frac{m}{\sin \theta} \,  v'_{\varphi} \right] \, - \, D_{\varphi \mu} \, \frac{1}{\sin \theta} \, \frac{\partial}{\partial \theta} (\sin \theta  v'_{\varphi}) \, \nonumber \\
&- \,  i { D}_{\zeta} v'_{\zeta} \, - \, i \sin \theta { D}_{\mu}  v'_{\theta} \, - \, \sin \theta  D_{\varphi}  v'_{\varphi} \, - \, \frac{1}{ r} \,  L^{2} \pi' \, - \, \frac{1}{ r} \,  L^{2} \Phi' \nonumber \\
&+ \,  D_{\Phi \mu} \frac{\partial \Phi'}{\partial \mu} \, + \,  D_{\rho} \rho'\, + \,  D_{\rho \mu} \frac{\partial \rho'}{\partial \mu} \, + \,  D_{\pi} \pi'\, + \,  D_{\pi \mu} \frac{\partial \pi'}{\partial \mu}\, ,  \\
\hbox{with,  }& \nonumber \\
&  D_{\varphi \mu} \, = \, \Omega \,t \nonumber \\
\label{def_t} &\hspace{1cm}\hbox{where} \hspace{1cm} t(\zeta,\mu) \, = \, 2 \left( \mu - (1-\mu²) \frac{ r_{\mu}}{ r} \right)\, , \\
&  D_{\zeta} = -\, m  g \, \left[ 2 \Omega \,  \left( 1  - \frac{ r_{\theta}}{ r} \, \cot \theta \right) \,+ \, \frac{ r}{ r_{\zeta}} (1 - \frac{ r_{\theta}^2}{ r^2})\, \frac{\partial \Omega}{\partial \zeta} \, - \, \frac{ r_{\theta}}{ r} \, \frac{\partial \Omega}{\partial \theta} \right] \, ,\nonumber \\
&  D_{\mu}\, = \, m \, \left[ \frac{\partial \Omega}{\partial \mu }(1+ g) \, - \, \frac{\Omega \,  g \,  t }{1-\mu^{2}} \right] \, ,\nonumber\\
&  D_{\varphi} \, = \, - \frac{\partial (\Omega \,  t)}{\partial \mu} \, , \nonumber \\
&  D_{\rho} \, = \, - \, \frac{\partial }{\partial \mu} \left( \frac{1-\mu^{2}}{ r \rho^{2}} \frac{\partial  p}{\partial \mu} \right)\, , \nonumber\\
&  D_{\rho \mu} \, = \, - \, \frac{(1-\mu^{2})}{ r \rho^{2}} \frac{\partial  p}{\partial \mu}\, , \nonumber \\
&  D_{\pi} \, = \, \frac{\partial }{\partial \mu} \left( \frac{1-\mu^2}{ r \rho} \frac{\partial \rho}{\partial \mu}\right) \, ,\nonumber\\
&  D_{\pi \mu} = \frac{1-\mu^{2}}{r} \left( \frac{1}{\rho} \frac{\partial \rho }{\partial \mu}- \frac{r_{\mu}}{r}  \right)\, , \nonumber \\
&  D_{\Phi \mu}\, = \, - \, \frac{(1-\mu^{2})  r_{\mu}}{ r^{2}} \, .\nonumber
\end{align}

\subsection{Toroidal motion}
		
The curl of the motion equation is \citep{Unno1989}
\begin{equation}
\left[ \left( \frac{\partial}{\partial t} + \Omega \frac{\partial}{\partial \phi} \right) \omega'_{i} \right] \mathbf{e_{i}}+\left( \mathbf{\omega'} . \mathbf{\nabla \Omega} \right)r \sin\theta \mathbf{e_{\phi}} + \left( \mathbf{v'}. \mathbf{\nabla} \right) 2 \mathbf{\Omega} - \left( 2 \mathbf{\Omega}.\mathbf{\nabla} \right) \mathbf{v'} + \left(\mathbf{\nabla}.\mathbf{v'} \right) 2 \mathbf{\Omega} + \mathbf{\nabla} \times \left(\frac{1}{\rho} \mathbf{\nabla}p \right) = \mathbf{0}\, ,
\end{equation}
where we introduce the vorticity vector: $\vec{\omega}\, = \, \vec{\nabla} \times \vec{v} $. In order to find the toroidal component of the motion, we take the pseudo-radial component of the above equation (\textit{i.e.} along $\overrightarrow{a_{\zeta}}$):
\begin{align}
\label{App_rot} 0&\, = \,  i(m\Omega +\sigma) \, \left[ \frac{1}{\sin \theta} \frac{\partial}{\partial \theta} (\sin \theta  v'_{\varphi}) \, - \, \frac{1}{\sin \theta}  \frac{\partial  v'_{\theta}}{\partial \varphi} \right] \, + \,  i (m \Omega+ \sigma) \frac{ r_{\theta}}{ r}  v'_{\varphi} \, + \,  i \sin \theta   R_{\varphi} \,  v'_{\varphi} \nonumber \\
&+ \,  R_{\zeta}\,  v'_{\zeta} \, + \, \sin \theta  R_{\mu}  v'_{\theta} \, + \, R_{\zeta \mu} \, \frac{\partial  v'_{\zeta}}{\partial \mu} \, + \,  R_{\mu \mu} \frac{1}{\sin \theta} \, \frac{\partial}{\partial \theta}\, (\sin \theta  v'_{\theta}) \, + \,  i  R_{\pi} \, \pi'\, + \,  i  R_{\rho} \, \rho' \\
\hbox{where, }& \nonumber \\
&\hspace{2.5cm}  R_{\zeta} \, = \,  a\,  t \, + \, (1-\mu^{2}) \, \frac{ r_{\mu}}{ r} \,  c \, + \, \Omega \,  t\, ( d+ f) \nonumber \\
&\hspace{2.5cm}  R_{\mu}\, = \, - \,  t \,  b \, + \,  c \, - \, 2\Omega \, + \, \Omega ( e+ h) \,  t \nonumber \\
&\hspace{2.5cm}  R_{\zeta \mu}\, = \, - \, \Omega (1-\mu^{2}) \, \left( 2+ \frac{ r_{\mu}}{ r} \,  g\,  t \right) \nonumber \\
&\hspace{2.5cm}  R_{\mu \mu}\, = \, \Omega \,  t \,  g \nonumber \\
&\hspace{2.5cm}  R_{\varphi} \, = \, \frac{m\Omega t}{(1-\mu^{2})} \nonumber \\
&\hspace{2.5cm}  R_{\pi} \, = \, - \, \frac{m}{ r \rho} \, \frac{\partial \rho}{\partial \mu} \nonumber \\
&\hspace{2.5cm}   R_{\rho} \, = \, \frac{m}{ r \rho^{2}} \frac{\partial  p}{\partial \mu} \nonumber
\end{align}
$t$ is given in Eq. (\ref{def_t}) and $a$, $b$, $c$, $d$, $e$, $f$ and $h$ are defined by:
\begin{align}
 a \, &= \, \frac{ r}{ r_{\zeta}}\, \frac{\partial \Omega}{\partial \zeta} \, - \, (1-\mu^2) \, \frac{ r_{\mu}}{ r} \, \frac{\partial \Omega}{\partial \mu} \nonumber \\
 b \, &= \,  \frac{\partial \Omega}{\partial \mu}\nonumber \\
 c \, &= \, \Omega \, \left( \frac{ t}{ r_{\zeta}} \, \frac{\partial  r_{\mu}}{\partial \zeta} \, + \, \frac{2 \, (1-\mu^2)}{ r} \frac{\partial  r_{\mu}}{\partial \mu} \, - \, 2 \, \mu \frac{ r_{\mu}}{ r}\right)\nonumber \\
 d \, &= \, 2 \, - \, (1-\mu^2)\, \frac{ r_{\mu}}{ r} \, \left( - \, \frac{ r_{\mu}}{ r} \, + \, \frac{1}{ r_{\zeta}} \, \frac{\partial  r_{\mu}}{\partial \zeta} \, + \, \mu \, \frac{ r_{\mu}}{ r}\right)\nonumber \\
 e \, &= \,  \frac{1}{ r_{\zeta}} \, \frac{\partial  r_{\mu}}{\partial \zeta} \nonumber \\
 f \, &= \, \frac{(1-2 g)}{ r} \, \left( \frac{1}{2} \,  r_{\mu}  t  \, + \, (1-\mu^2) \, \frac{\partial  r_{\mu}}{\partial \mu} \, - \, 2 \, \mu \,  r_{\mu}\right)\nonumber \\
 h \, &= \, 2 \, \frac{ r_{\mu}}{ r} \,  g \, \left[ -(1-\mu^2) \, \left( \frac{1}{ r} \, \frac{\partial  r_{\mu}}{\partial \mu} \, - \, \frac{ r_{\mu}^2}{ r^2} \right)\, + \, \mu \, \frac{ r_{\mu}}{ r} \right]\nonumber
\end{align}

\subsection{Adiabatic relation}

Let us define quantities analogous to the Schwarzschild discriminant in the two dimensional case:

\begin{align}
 A_{\zeta} \, &= \, \frac{1}{\Gamma_{1}} \left( \frac{\partial \ln  p}{\partial \zeta} \right)_{\mu} \, - \, \left( \frac{\partial \ln \rho}{\partial \zeta}\right)_{\mu} \\
 A_{\mu} \, &= \, \frac{1}{\Gamma_{1}} \left(\frac{\partial \ln  p}{\partial \mu}\right)_{\zeta} \, - \, \left(\frac{\partial \ln \rho}{\partial \mu} \right)_{\zeta}
\label{App-Eq_defAtild}
\end{align}
Therefore, the adiabatic relation given in Eq. (\ref{relat_ad_pert}) can be written under the following form:

 \begin{align}
 \label{App_eq_relat_adiab2}
  i( m \Omega+\sigma) \, \left( \frac{1}{\rho} \, \rho'\, - \, \frac{\rho}{\Gamma_{1} \,  p} \, \pi' \right) \, - \, \tilde{ A}_{\zeta}  v'_{\zeta} \, + \, \sin \theta \, \tilde{ A}_{\mu} \, \, v'_{\theta}=0\, ,
 \end{align}
where we have introduced: $\tilde{ A}_{\mu}\, =\,  g \,  A_{\mu} \, /  r  $, and $\tilde{ A}_{\zeta} \, = \,  A_{\zeta}/  r_{\zeta} \, - \, (1-\mu^2) \,  r_{\mu}\, \tilde{ A}_{\mu}/  r $.

\subsection{Conservation of mass}

After linearization of Eq. (\ref{conservation_pert_xi}), one obtains:

\begin{equation}
 i(m\Omega+\sigma) \, \frac{\rho'}{\rho} \, + \, \mathbf{\nabla} \cdot \mathbf{v'} \, + \, \mathbf{v'} \cdot \mathbf{\nabla} \ln \rho \, = \, 0  \nonumber 
\end{equation}

With the help of Eq. (\ref{relat_ad_pert}), the equation can be rewritten:  
	
\begin{equation}
 i(m\Omega+\sigma) \, \frac{\rho}{\Gamma_{1} \,  p} \, \pi'\, + \, \frac{1}{\Gamma_{1}} \, \mathbf{v'} \cdot \mathbf{\nabla} \ln  p \, + \,  \mathbf{\nabla}.\mathbf{v'} \, = \, 0 \nonumber 
\end{equation}
which finally gives:

\begin{align}
\label{App_eq_Continuite}
\frac{\partial  v'_{\zeta}}{\partial \zeta} \, &= \, - \,  i( m\Omega+\sigma) \,  \frac{ r_{\zeta}\rho}{\Gamma_{1}  p} \, \pi' \, - \,  g \,  \frac{ r_{\zeta}}{ r \sin \theta} \, \frac{\partial }{\partial \theta} ( \sin \theta  v'_{\theta}) \, - \, \frac{ r_{\zeta}}{ r \sin \theta}  \frac{\partial  v'_{\varphi}}{\partial \varphi} \nonumber \\
&+ \,  C_{\zeta} \,  v'_{\zeta} - \sin \theta \,  C_{\mu} v'_{\theta} \, + \,  C_{\zeta,\mu} \, \frac{\partial  v'_{\zeta}}{\partial \mu} \\
\hbox{with, }& \nonumber \\
& C_{\zeta}\, = \, \frac{1}{\Gamma_{1} \,  p} \left(  g \, \frac{(1-\mu^{2})  r_{\mu}  r_{\zeta}}{ r^{2}} \frac{\partial  p}{\partial \mu}  \, - \, \frac{\partial  p}{\partial \zeta} \right) \, + \,  g \, \frac{ r_{\zeta}}{ r} ((1-\mu^2) \, \frac{ r_{\mu}}{2  r} \, - \,\frac{ r}{2  r_{\mu}} \,  h \, - \,  d ) \nonumber \\
&  C_{\mu} \, = \, - \,  g \, \frac{ r_{\zeta}}{ r} \, \left( \frac{1}{\Gamma_{1} \,  p} \, \frac{\partial  p}{\partial \mu} \, + \,  e \, + \,  h \, + \, \frac{ r_{\mu}}{ r} \right) \nonumber \\
&  C_{\zeta \mu} \, =  \, (1-\mu^{2}) \,   g \, \frac{ r_{\zeta}r_{\mu}}{ r^{2}}  \nonumber 
\end{align}	

\subsection{Poisson's equation}

Computing the Laplacien of $\Phi'$ in the new coordinate system, one obtains a dimensionless version of Poisson's equation:

\begin{align}
\rho' =& \frac{1}{r_{\zeta}^{2}.g(\zeta,\mu)} \frac{\partial^{2} \Phi'}{\partial \zeta^{2}}+\frac{1}{r_{\zeta}} \frac{\partial \Phi'}{\partial \zeta} \left[ \frac{r_{\theta}}{r^{2}} \frac{\partial }{\partial \zeta} \left( \frac{r_{\theta}}{r_{\zeta}} \right)+ \frac{1}{r^{2}} \frac{\partial }{\partial \zeta} \left(\frac{r^{2}}{r_{\zeta}} \right) - \frac{r_{\zeta}}{r^{2} \sin\theta} \frac{\partial}{\partial \theta}(\frac{r_{\theta}}{r_{\zeta}} \sin\theta) \right] \nonumber \\
+&\frac{1}{r^{2}\sin \theta} \frac{\partial}{\partial \theta} \left( \sin \theta \frac{\partial \Phi'}{\partial \theta} \right) - \frac{2 r_{\theta}}{r_{\zeta}r^{2}} \frac{\partial^{2} \Phi'}{\partial \zeta \partial \theta} + \frac{1}{r^{2}\sin^2 \theta} \frac{\partial^{2} \Phi'}{\partial \phi^{2}}
\end{align}
As mentioned in Sect. \ref{ss_basic-eq}, we introduce a new variable and add an auxiliary equation in order to have a first order differential system:
\begin{equation}
d\Phi'= \partial_{\zeta} \Phi'
\label{App_eq-aux}
\end{equation}

\begin{align}
\label{App_eq_Poisson2}
\frac{\partial  d\Phi'}{\partial \zeta} \, &= \,  P_{\rho'} \, \rho' \, + \,  P_{d\Phi'} \,  d\Phi' \, + \,  P_{d\Phi' \mu} \, \frac{\partial   d\Phi'}{\partial \mu} \, + \,  P_{\Phi' \mu}  L^2(\Phi')\\
\hbox{with, } & \nonumber \\
&  P_{d\Phi'} \, = \, \frac{ g}{ r_{\zeta}} \, \frac{\partial  r_{\zeta}}{\partial \zeta} \, - \,  g \, \frac{ r_{\zeta}}{ r} \, (2+2\mu \frac{ r_{\mu}}{ r}-(1-\mu^2) \frac{1}{ r} \frac{\partial  r_{\mu}}{\partial \mu}) \nonumber \\
& P_{d\Phi' \mu} \, = \,2(1-\mu^{2}) \, \frac{ r_{\mu}r_{\zeta}}{ r^{2}} \,  g \nonumber \\
&  P_{\phi' \mu} \, = \,\frac{ r_{\zeta}^{2}}{ r^{2}} \,  g \hspace{2cm} \nonumber \\
\label{App_eq_P_dfimu}
& P_{\rho'} \, = \,  r_{\zeta}^{2} \,  g 
\end{align} 

Eqs.~(\ref{App_eq_Mouvement}), (\ref{App_eq_Divergence}), (\ref{App_rot}), (\ref{App_eq_relat_adiab2}), (\ref{App_eq_Continuite}), (\ref{App_eq-aux}) and~(\ref{App_eq_Poisson2}) are the 7 equations which make up the first-order differential equation system which yields 2D pulsations of any rotating model.

In the second domain $V_2$, $\mathrm{i.e.}$ in the vacuum, only Eqs.~(\ref{App_eq-aux}) and~(\ref{App_eq_Poisson2}) remain with $\rho'=0$ for Poisson's equation.

\section{Central boundary conditions}
\label{app_cond-cent}

As explained in Sect. \ref{bound_cond}, we choose to impose two boundary conditions at the center.
The goal of this section is to find two algebraic equations to replace differential ones at the center of the star.

Table \ref{C4_tab_term} shows the central behavior of the first terms in the spectral expansion of the unknowns. As can be seen in the table, the parity of the $w_{\ell}$ coefficients is different from the parity of other spectral coefficients,

\begin{table}[h]
\caption{\label{C4_tab_term} Central behavior of the first terms in the spectral expansion in different cases: in blue are given the \textcolor{blue}{odd modes} and in red the \textcolor{red}{even modes}.
}
\vspace{0.3cm}
\centering
\begin{tabular}{lccc}
\hline \hline
\vspace{-0.1cm}\\
 & $u_{\ell}, \, v_{\ell}$ & $w_{\ell}$ & $\pi'_{\ell}, \, \Phi'_{\ell}, \, \rm d\Phi'_{\ell}, \, \rho'_{\ell} $ \vspace{0.3cm}   \\ 
\hline
\vspace{-0.2cm}\\
\color{blue}{$m \, =  \, 0\, $ and $ \,  {par} \, =  \, 1 $} & $u_1\, \propto \, \zeta^0 \, $ & $w_2 \, \propto \, \zeta^2$ & $\pi'_1 \, \propto \, \zeta$ \\
& $u_3\, \propto \, \zeta^2 \, $ & $w_4 \, \propto \, \zeta^4$ & $\pi'_3 \, \propto \, \zeta^3$ \vspace{0.2cm} \\
\hline
\vspace{-0.2cm}\\
\color{red}{$m \, \ne  \, 0 \,$ and $ \,  {par} \, =  \, 0 $} & $u_{\mid m \mid}\, \propto \, \zeta^{\mid m \mid -1} \, $ & $w_{\mid m \mid +1} \, \propto \, \zeta^{\mid m \mid +1}$ & $\pi'_{\mid m \mid} \, \propto \, \zeta^{\mid m \mid}$ \\
& $u_{\mid m \mid+2}\, \propto \, \zeta^{\mid m \mid +1} \, $ & $w_{\mid m \mid +3} \, \propto \, \zeta^{\mid m \mid +3}$ & $\pi'_{\mid m \mid+2} \, \propto \, \zeta^{\mid m \mid+2}$ \vspace{0.2cm} \\
\hline
 \vspace{-0.2cm}\\
\color{red}{$m \, =  \, 0 \,$ and $ \,  {par} \, =  \, 0 $} & $u_{0}\, \propto \, \zeta \,$, $v_{0} \equiv 0$  & $w_1 \, \propto \, \zeta$ & $\pi'_0 \, \propto \, \zeta^0$\\
& $u_{2}\, \propto \, \zeta \, $ & $w_{3} \, \propto \, \zeta^{3}$ & $\pi'_{2} \, \propto \, \zeta^{2}$ \vspace{0.2cm} \\
\hline
 \vspace{-0.2cm}\\
\color{blue}{$m \, \ne  \, 0 \, $ and $\,  {par} \, =  \, 1 $} & $u_{\mid m \mid+1}\, \propto \, \zeta^{\mid m \mid} \, $ & $w_{\mid m \mid} \, \propto \, \zeta^{\mid m \mid}$ & $\pi'_{\mid m \mid+1} \, \propto \, \zeta^{\mid m \mid+1}$\\
& $u_{\mid m \mid+3}\, \propto \, \zeta^{\mid m \mid +2} \, $ & $w_{\mid m \mid +2} \, \propto \, \zeta^{\mid m \mid +2}$ & $\pi'_{\mid m \mid+3} \, \propto \, \zeta^{\mid m \mid+3}$ \vspace{0.2cm} \\
\hline
\end{tabular}
\end{table}

The global parity of the mode is defined by the parity of  $u_{\ell}, \, v_{\ell}$, $\pi'_{\ell}, \, \Phi'_{\ell}, \, \rm d\Phi'_{\ell}$ and $\rho'_{\ell} $. 
Note that in the case of an odd axisymmetric mode, $w_{0}$ has no meaning since the spherical harmonic $Y_{\ell=0,\,m=0}$ is spherically symmetric and has no toroidal component.  A similar argument applies to the $v_{0}$ component in even axisymmetric modes.

In order to emphasize the dominant terms in the equations near the center, we propose to express the spectral components under the following form: 

\begin{align}
&\pi'_{\ell}(\zeta) \, = \, \zeta^{\ell} \, \tilde{\pi_{\ell}}'(\zeta)  \hspace{1.4cm} \Phi'_{\ell}(\zeta) \, = \, \zeta^{\ell} \, \tilde{\Phi_{\ell}}'(\zeta)  \hspace{1.2cm}  \rho'_{\ell}(\zeta) \, = \, \zeta^{\ell} \, \tilde{\rho_{\ell}}'(\zeta) \nonumber \\
&u_{\ell}(\zeta) \, = \, \zeta^{\ell-1} \, \tilde{u_{\ell}}(\zeta) \hspace{1.1cm} v_{\ell}(\zeta) = \zeta^{\ell-1} \, \tilde{v_{\ell}}(\zeta) \hspace{1.2cm} w_{\ell}(\zeta) \, = \, \zeta^{\ell} \, \tilde{w_{\ell}}(\zeta)
\label{App2_eq_dev_zetal}
\end{align}
where the tilded quantities have a constant behavior at the center. For the derivative of the gravitational potential, we obtain: $d\Phi'_{\ell} = \zeta^{\ell+1}\,\tilde{d\Phi'_{\ell}} \, +\, \ell \, \zeta^{\ell-1}\, \tilde{\Phi'_{\ell}}$.

\noindent Introducing this variable change into the system of equations, and taking the limit when $\zeta$ goes to zero, we obtain:

\vspace*{0.5cm}\begin{center} \boxed{\hbox{General case:}} \end{center} 

\begin{align}
&\hbox{Radial motion (Eq. \ref{App_eq_Mouvement})} \nonumber \\
&\zeta^{\ell} \, \frac{\partial \tilde{\pi}_{\ell}'}{\partial \zeta} \, = \, \Big[ (m \Omega + \sigma) \, (1-\epsilon)\, \tilde{u}_{\ell}'\, - \, 2 \, m \Omega \, (1-\epsilon) \,  \tilde{v}_{\ell}'\, - \, \ell \, \tilde{\Phi}'_{\ell} \, - \, \ell \, \tilde{\pi}'_{\ell} -\, 2 \,(\ell-1) \, { J}_{\ell}^{m} \, \Omega \, (1-\epsilon) \,  \tilde{w}_{\ell-1}'\, \Big] \, \zeta^{\ell-1} \nonumber \\
\label{App2_Eq_Mouvcentlm}
&\hspace{1.2cm} + \left[ \, \mathcal{M}_{\rho} \, \tilde{\rho}_{\ell}'\, + \, \mathcal{M}_{\pi} \, \tilde{\pi}_{\ell}' \, + \,  2 \,(\ell+2) \, { J}_{\ell+1}^{m} \, \Omega \, (1-\epsilon) \,  \tilde{w}_{\ell+1}'\,- \, \rm d\tilde{\Phi}'_{\ell} \right] \, \zeta^{\ell+1}\, \\
&\hbox{Poloidal motion (Eq. \ref{App_eq_Divergence})} \nonumber \\
&0 \, = \, \Big[ \left( (m \Omega + \sigma) \, \, \ell (\ell+1) \, -\, 2 m \Omega \right) \tilde{v}_{\ell}' \, - \, 2 m \,  \Omega \,  \tilde{u}_{\ell}' \, - \, 2\,  \Omega \, (\ell+1)\, (\ell -1) \,  J_{\ell}^{m} \,  \tilde{w}_{\ell-1}'\, \nonumber \\
\label{App2_Eq_Divcentlm}
& \hspace{0.8cm}- \, \frac{\zeta}{ r} \, \ell (\ell+1)\,  \tilde{\pi}_{\ell}' \,- \, \frac{\zeta}{ r} \, \ell (\ell+1)\, \tilde{\Phi}_{\ell}' \, \Big] \, \zeta^{\ell-1} \, - \, 2\,  \Omega \,  \ell \,  (\ell +2) \,  J_{\ell+1}^{m} \,  \tilde{w}_{\ell+1}' \, \zeta^{\ell+1}  \\
&\hbox{Toroidal motion (Eq. \ref{App_rot})} \nonumber \\ 
\label{App2_Eq_Rotcentlm}
&0 \, =  \, 2 \Omega \, \left[ - \, (\ell_{1p}-1)\, (\ell_{1p}+1)\,  J_{\ell_{1p}}^{m} \,\tilde{v}_{\ell_{1p}-1}' \, + \, (\ell_{1p}+1) \, J_{\ell_{1p}}^{m} \, \tilde{u}_{\ell_{1p}-1}' \right] \, \zeta^{\ell_{1p-2}}  \\
&\hspace{0.3cm} + \, \left[  \left(  (m \Omega + \sigma) \ell_{1p}(\ell_{1p}+1)-2 m \Omega \right) \, \tilde{w}_{\ell_{1p}}' \, -  \, 2 \Omega \, \ell_{1p}\, J_{\ell_{1p}+1}^{m}  \left( (\ell_{1p}+2) \, \tilde{v}_{\ell_{1p}+1}' \, +\, \tilde{u}_{\ell_{1p}+1}' \right) \right] \zeta^{\ell_{1p}}    \nonumber \\
&\hbox{Adiabatic relation (Eq.\ref{App_eq_relat_adiab2})} \nonumber \\ 
\label{App2_Eq_Adcentlm}
 &0 \, = \,  (m \Omega + \sigma) \frac{1}{\rho_0} \zeta^{\ell} \tilde{\rho}_{\ell}' \, - \,(m \Omega + \sigma) \frac{\rho_0}{\Gamma_1  P_0} \zeta^{\ell} \tilde{\pi}_{\ell}' \, - \, \frac{1}{ r_{\zeta}} \, \mathcal{A}_{\zeta} \, \zeta^{\ell} \, \tilde{u}_{\ell}'\\
&\hbox{Continuity equation (Eq. \ref{App_eq_Continuite})} \nonumber \\ 
\label{App2_Eq_Contcentlm}
& \zeta^{\ell-1} \, \frac{\partial \tilde{u}_{\ell}'}{\partial \zeta} \, = \, \left[  \left( -(\ell-1) \, + \, \mathcal{C}_{\zeta}\right)  \,   \tilde{u}_{\ell}' \, + \,  \ell(\ell+1) \,  \tilde{v}_{\ell}' \right] \, \zeta^{\ell-2} \, - \, (1- \epsilon) (m \Omega + \sigma) \frac{\rho_0}{\Gamma_1  P_0} \zeta^{\ell} \, \tilde{\pi}_{\ell}' \\
&\hbox{Poisson's equation (Eq. \ref{App_eq_Poisson2})} \nonumber \\ 
\label{App2_Eq_Poiscentlm}
& \zeta^{\ell+1} \, \frac{\partial {\rm d}\tilde{\Phi}_{\ell}'}{\partial \zeta} \, = \, (1-\epsilon)^{2}   \,\zeta^{\ell} \, \tilde{\rho}_{\ell}' \, - \, (2 \ell +3) \zeta^{\ell} \,  d\tilde{\Phi}_{\ell}' 
\end{align}
where $J_{\ell}^m$ is given by:
\begin{align}
J_{\ell}^m \, = \, \begin{cases} \sqrt{\frac{\ell^2-m^2}{4 \ell^2 - 1}} \, ,& \, \hbox{ if }\, \ell > |m|; \\ 0  \, ,& \, \hbox{ if }\,  \ell \le |m|\, , \end{cases}
\end{align}
and where $\epsilon$ is the flatness, given in Sect. \ref{S2s1}.

We note, first of all, that the equations of radial motion, of continuity as well as Poisson's equation are singular at the center.
In order to enforce the regularity of those equations, the boundary conditions that are imposed, are given by the fact that singular terms go to zero when $\zeta \, \to \, 0$. Therefore the radial motion and Poisson's equations lead to

\begin{align}
\label{App2_Eq_Cond_mouv}
0 \, =& \,(m \Omega + \sigma) \, (1-\epsilon)\, \tilde{u}_{\ell}'\, - \, 2 \, m \Omega \, (1-\epsilon) \,  \tilde{v}_{\ell}'\, - \, \ell \, \tilde{\Phi}'_{\ell} \, - \, \ell \, \tilde{\pi}'_{\ell} -\, 2 \,(\ell-1) \, { J}_{\ell}^{m} \, \Omega \, (1-\epsilon) \,  \tilde{w}_{\ell-1}' \\
0 \, =& \, (1-\epsilon)^{2}   \,\tilde{\rho}_{\ell}' \, - \, (2 \ell +3) \,  d\tilde{\Phi}_{\ell}' 
\label{App2_Eq_Cond_pois}
\end{align}

\vspace*{0.5cm}\begin{center} \boxed{\hbox{Case $\mid m \mid = 0$ and $par=0$ and $\ell=0$:}} \end{center} 

This case is specific in the sense that, near the center, $u_{0} \, = \, \mathcal{O} \left( \zeta \right)$. Thus, the radial motion equation is not singular, whereas the continuity equation is:

\begin{align}
 \zeta \, \frac{\partial \tilde{u}_{0}'}{\partial \zeta} \, = \, \left[ \, - \, (1-\epsilon) \, \sigma \,  \frac{\rho_0}{\Gamma_1  P_0} \tilde{\pi}_{0}'  \, +\,  \left(  \mathcal{C}_{\zeta} - 1 \right) \, \tilde{u}_{0}' \right] \, \zeta^0
\end{align}
Therefore, the algebraic equation to impose near the center when $\mid m \mid = 0$, $par=0$ and $\ell=0$ is:
\begin{align}
 0\, = \,  \, - \, (1-\epsilon) \, \sigma \,  \frac{\rho_0}{\Gamma_1  P_0} \tilde{\pi}_{0}'  \, +\,  \left(  \mathcal{C}_{\zeta} - 1 \right) \, \tilde{u}_{0}' 
\end{align}
Concerning Poisson's equation, it is also singular in this specific case, and Eq. (\ref{App2_Eq_Cond_pois}) is still relevant.
 
It seems then that we obtained the algebraic boundary conditions necessary to impose regular behavior of the unknowns near the center. However, the condition given in Eq. (\ref{App2_Eq_Cond_mouv}) is redundant with a linear combination of Eqs.~(\ref{App2_Eq_Divcentlm}) and~(\ref{App2_Eq_Rotcentlm}). Hence, the system is under-determined at the center. 
It is not possible to enforce the right behavior to the unknowns near the center using the tilded variables.
We therefore have to impose central conditions on the non-tilded spectral components directly. Here again, different cases have to be explored:

\vspace*{0.5cm}\begin{center} \boxed{\hbox{General case:}} \end{center} 
The general case is composed of the following symmetry cases:
\begin{itemize} 
\item $\mid m \mid = 0$ and $par=1$, 
\item $\mid m \mid \ne 0$ and $par=0$,\\
\end{itemize}
as well as all the components after the first one for: 
\begin{itemize}
\item $\mid m \mid = 0$ and $par=0$, $\mathrm{i.e.}$ $\ell=2,4,6,\cdots$
\item $\mid m \mid \ne 0$ and $par=1$, $\mathrm{i.e.}$ $\ell=\mid m \mid +3, \mid m \mid +5, \mid m \mid +7, \cdots$
\end{itemize}

\begin{align}
\label{C4_Cond_centre1}
\hbox{The regularity of the velocity field leads to }\hspace{0.75cm}0 \, &= \, u_{\ell} \, - \, \ell \, v_{\ell} \\
0 \, &= \, w_{\ell}  \\
\hbox{The regularity of the motion equation leads to }\hspace{0.2cm}0\,  &= \, (1-\epsilon)(m\Omega+\sigma) u_{\ell}\, - \, 2m\Omega (1-\epsilon) v_{\ell}\, - \, \frac{\ell}{\zeta} \Phi'_{\ell}\, - \, \frac{\ell}{\zeta} \pi'_{\ell} \\
\hbox{The regularity of Poisson's equation leads to }\hspace{0.45cm}0 \, &= \,\rm d\Phi'_{\ell} \, - \, \frac{\ell}{\zeta}\,\Phi'_{\ell}\, 
\end{align}

\vspace*{0.5cm}\begin{center} \boxed{\hbox{Case $\mid m \mid = 0$, $par=0$ and $\ell=0$:}} \end{center} 
In this specific case, $w_{\ell-1}$ has no meaning, we thus impose
\begin{align}
0 \, &= \, v_{0} \, ,\\
0\,  &= \, (1-\epsilon) \, \zeta \, \sigma^2 \, \frac{\rho_0}{\Gamma_1 \,  P_0} \, \pi'_{0}\, + \, 3 \, u_{0}  \hspace{3cm}\, , \\
0 \, &= \,\rm d\Phi'_{0} \, ,
\end{align}
where the condition over the radial velocity component is obtained from the continuity equation.

\vspace*{0.5cm}\begin{center} \boxed{\hbox{Case $\mid m \mid \ne 0$, $par=1$ and $\ell=\mid m \mid+1$:}} \end{center} 
The continuity equation, the toroidal and radial motion equations, and Poisson's equation give respectively 
\begin{align}
0 \, &= \, u_{\mid m \mid+1} \, - \, \ell \, v_{\mid m \mid+1} \, , \\
0 \, &= \, \left(  (m \Omega + \sigma) \mid m \mid \, (\mid m \mid+1)-2 m \Omega \right) \, w_{\mid m \mid}' \, , \nonumber  \\
&\hspace{0.2cm} -  \, 2 \Omega \, \mid m \mid \,{ J}_{\mid m \mid+1}^{m}  \left( (\mid m \mid+2) \, v_{\mid m \mid+1}' \, +\, u_{\mid m \mid+1}' \right)  \, , \\
0 \, &= \,(m \Omega + \sigma) \, (1-\epsilon)\, u_{\mid m \mid+1}'\, - \, 2 \, m \Omega \, (1-\epsilon) \,  v_{\mid m \mid+1}'\, - \, \ell \, \Phi'_{\mid m \mid+1}  \, , \nonumber \\
&\hspace{0.2cm} - \, \ell \, \pi'_{\mid m \mid+1} -\, 2 \,\mid m \mid \, { J}_{\mid m \mid+1}^{m} \, \Omega \, (1-\epsilon) \,  w_{\mid m \mid}' \, ,  \\
0 \, &= \, {\rm d}\Phi'_{\mid m \mid+1} \, -\, \frac{\mid m \mid+1}{\zeta}\,\Phi'_{\mid m \mid+1}\, .
\label{C4_Cond_centre2}
\end{align}

The components $w_{\ell}$ are neglected at the center in the remaining cases. This is an approximation, but after verifications, this method is the most efficient and numerically suitable in order to enforce regular behavior of the solutions at the center.

\section{Comparison of eigenfunctions}

\label{App_comp_modes}
\vspace*{1.5cm}
 \begin{figure} [h!]
\begin{center}
\includegraphics[scale=0.9]{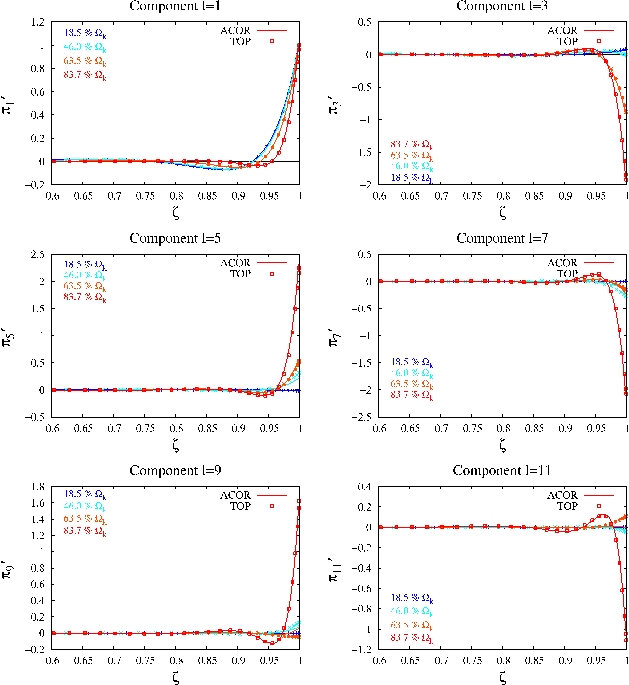}
\caption{\label{fig_Pi-l_10_bf2} Radial parts of the eigenfunction  $\pi'_{\ell}$ for a low frequency mode dominated by an $\ell=1$, $m=0$ component, for an $N=3$ polytropic model uniformly rotating at $18.5 \%$, $46 \%$, $63.5 \%$ and $83.7 \%$ of the Keplerian break-up velocity.  
}
\end{center}
\end{figure}

\begin{figure} [p]
\vspace*{2.8cm}
\begin{center}
\includegraphics[scale=0.9]{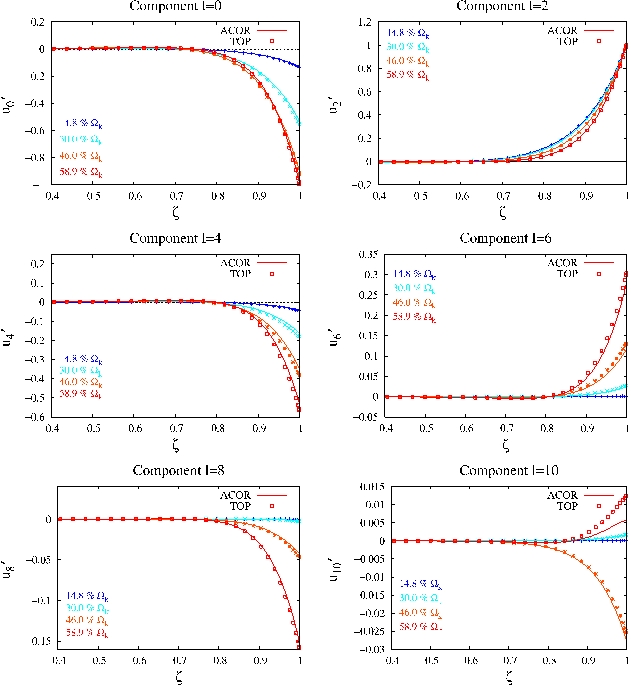}
\caption{\label{fig_Pi-l_20_bf2} Radial parts of the eigenfunction  $\pi'_{\ell}$ for a low frequency mode dominated by an $\ell=2$, $m=0$ component, for an $N=3$ polytropic model uniformly rotating at $3.7 \%$, $18.5 \%$, $46 \%$ and $64.5 \%$ of the Keplerian break-up velocity.  
}
\end{center}
\end{figure}
\end{appendix}
\end{onecolumn}

\end{document}